%% file: main.tex
\documentclass[preprint]{elsarticle}

\journal{Journal of Parallel and Distributed Computing}

\renewcommand*{\today}{November 5, 2019}

\usepackage{xcolor}
\usepackage{lipsum}
\usepackage{hyperref}
\usepackage{amstext}
\usepackage{booktabs}
\usepackage{subfig}
\usepackage{tikz}
\usepackage{pgfplots}
\usepackage{siunitx}

\pgfplotsset{compat=1.3}

\usepackage{listings}
\usepackage{float}
\newfloat{lstfloat}{H}{lop} 

\lstdefinestyle{customFortran1}{
  basicstyle=\ttfamily \scriptsize,
  keywordstyle=\bf,
  morekeywords={then,end,type,integer,use,call,logical},
  commentstyle=\itshape \color{gray},
  morecomment=[l]{!},
  frame=tb,
  numbers=left,
  numberstyle=\scriptsize,
  numbersep=2pt,
  stepnumber=1,
  tabsize=2,
  basewidth={0.55em},
  keepspaces,
  columns=fixed
}

\lstdefinestyle{customFortran2}{
  basicstyle=\ttfamily \footnotesize,
  keywordstyle=\bf,
  morekeywords={then,end,type,integer,use,call,logical},
  commentstyle=\itshape \color{gray},
  morecomment=[l]{!},
  frame=none,
  numbers=none,
  basewidth={0.55em},
  keepspaces,
  columns=fixed
}

\newcommand{\pmf}[1]{{\color{black}{#1}}}    
\newcommand{\rev}[1]{{\color{black}{#1}}}    
\newcommand{\vb}[1]{{\color{black}{#1}}}    
\newcommand{\xt}[1]{{\color{black}{#1}}}   
\newcommand{\jc}[1]{{\color{black}{#1}}} 
\newcommand{\rp}[1]{{\color{black}{#1}}}    

\begin{document}

\begin{frontmatter}

\title{HDOT - an Approach Towards Productive Programming of Hybrid Applications}


\author[mymainaddress]{Jan Ciesko}

\author[mymainaddress]{Pedro~J.~Mart\'inez-Ferrer\corref{mycorrespondingauthor}}
\cortext[mycorrespondingauthor]{Corresponding author}
\ead{pedro.martinez-ferrer@bsc.es}

\author[mymainaddress]{Ra\'ul Pe{\~n}acoba Veigas}

\author[mymainaddress]{Xavier Teruel}

\author[mymainaddress]{Vicen\c{c} Beltran}

\address[mymainaddress]%
{%
Barcelona Supercomputing Center (BSC)
}

\tnotetext[1]{\textcopyright 2019. This manuscript version is made available under the CC-BY-NC-ND 4.0 license \color{blue}{\url{https://creativecommons.org/licenses/by-nc-nd/4.0/}}}
\tnotetext[2]{Published journal article available at \color{blue}{\url{https://doi.org/10.1016/j.jpdc.2019.11.003}}}

\input{tex/abstract}


\end{frontmatter}


\input{tex/introduction}
\input{tex/background}
\input{tex/methodology}
\input{tex/heat}

\input{tex/HPCCG}
\input{tex/creams}
\input{tex/conclusion}

\section*{Acknowledgements}

This work has been developed with the support of the European Union
H2020 program through the INTERTWinE project (agreement number
671602); the Severo Ochoa Program awarded by the Spanish Government
(SEV-2015-0493); the Generalitat de Catalunya (contract
2017-SGR-1414); and the Spanish Ministry of Science and Innovation
(TIN2015-65316-P, Computaci\'on de Altas Prestaciones VII).  \rev{The
  authors gratefully acknowledge Dr.\ Arnaud Mura, CNRS researcher at
  Institut PPRIME in France, for the numerical tool CREAMS.}
\jc{Finally, the manuscript has \rev{greatly} benefited from the
  precise comments of the reviewers.}

\section*{References}
\bibliography{bib/main}

\end{document}

%% file: tex/abstract.tex
\begin{abstract}

\pmf{A wealth of important scientific and engineering applications are
  configured for use on high performance computing architectures using
  functionality found in the MPI specification.  This specification
  provides application developers with a straightforward means for
  implementing their ideas for execution on distributed-memory
  parallel processing computers. OpenMP directives provide a means for
  operating on shared-memory regions of those computers.  With the
  advent of machines composed of many-core processors, the strict
  synchronisation required by the bulk synchronous parallel (BSP)
  communication model can hinder performance increases.}  This is due
to the complexity to handle load imbalances, to reduce
serialisation imposed by \pmf{blocking} communication patterns, to
overlap communication with computation and, finally, to deal with
increasing memory overheads. The MPI specification provides advanced
features such as \pmf{non-blocking} calls or shared memory to mitigate
some of these factors.  However, applying these features efficiently
usually requires significant changes on the application structure.

\pmf{Task parallel programming models are being developed as a means
  of mitigating the abovementioned issues but without requiring
  extensive changes on the application code.  In this work, we present
  a methodology to develop hybrid applications based on tasks called
  \emph{hierarchical domain over-decomposition with tasking (HDOT)}.
  This methodology overcomes most of the issues found on MPI-only and
  traditional hybrid MPI+OpenMP applications}. However, by emphasising
the reuse of data partition schemes from process-level and applying
them to task-level, it enables a natural coexistence between MPI and
shared-memory programming models. \pmf{The proposed methodology} shows
promising results in terms of programmability and performance measured
on a set of applications.
\end{abstract}

%% file: tex/introduction.tex
\section{Introduction}\label{sec:intro}

Non-coherent, distributed memory is a common characteristic of modern
HPC systems. Such memory organisation allows to assemble large systems
from commodity components which reduces cost, increases versatility of
use and offers flexibility to quickly adapt to application
trends. This is referred to as cluster architecture.

MPI~\cite{mpi_standard} is a commonly used \pmf{programming
  specification} of such systems. In this scenario, programmers work
at the process level \pmf{(i.e.\ MPI rank)}, where each of these
processes has its \emph{memory address space} and therefore require
explicit communication for data exchange.  The programmer uses
\pmf{functions} such as \pmf{{\tt MPI\_SEND} or {\tt MPI\_RECV} to
  exchange and synchronise data between processes}. In other words,
the developer implements data distribution and coherence, such that
this functionality becomes part of the algorithm. Designing algorithms
with concurrency in mind makes software development more difficult,
which can be overwhelming for novice programmers. However, it turns
out that the nature of MPI, and imperative parallel programming models
in general, obliges the developer to consider concurrency early on in
the development. Typically this results in good design that favors
concurrency and scalability. As a consequence, many \pmf{MPI-only
  applications, i.e.\ those whose parallelism is implemented using
  functionality from the MPI specification}, tend to achieve better
performance scalability compared to applications using incremental
parallelism.

Algorithmic design for performance and scalability typically follows
one rule: \pmf{``always keep the processors busy advancing the
  computation''}. Under the hood this means: (i) implementing ordering
of computation and communication for overlaps, (ii) ensuring balanced
execution, (iii) keeping overheads low and, finally, (iv) creating
enough parallelism.  It turns out that this is becoming \pmf{a real
  challenge since the increasing} number of processor cores per
node\footnote{Currently, the Intel Skylake processor architecture
  supports up to 28 physical cores \rev{(the recently announced
    Cascade Lake processor will double this number) whilst the already
    available AMD EPYC 7742 CPU features 64 physical cores.}} makes it
more and more difficult to achieve scalability on these systems. This
is due to the fact that, as the number of processes within the same
node increases, efficiency of each MPI process
drops~\cite{Hoefler2012}.  The main causes for this behaviour are data
overheads, system heterogeneity, load imbalances and suboptimal
communication patterns, all of which are difficult to eliminate or
optimise.  \vb{In such cases, it can be beneficial to combine the
  message-passing programming model to exploit internode parallelism
  with another shared-memory programming model to exploit intranode
  parallelism~\cite{Yan2011, Barrett2015}.}

\subsection{Are shared-memory programming models the right solution?}

On shared memory programming models, threads are building blocks to
exploit parallelism within one \pmf{MPI} process. On a shared data
environment, load balancing techniques can be efficiently implemented,
there is no need to replicate extra data and all the threads have
direct access to all the data environment of their MPI
process. However, to benefit from threading, \pmf{an existing MPI-only
  code} must be adapted. A variety of ways exist to express
concurrency on thread-level of which OpenMP is a particularly popular
one (see Section~\ref{subsec:threading_mpi}). An application that
combines multiple programming models is called hybrid application.

\rev{In practice, h}ybrid codes can indeed offer an improvement over
\pmf{MPI-only} codes on modern many-core
systems~\cite{Chatterjee2013}.  However, it is up to the skill of the
programmer to combine both efficiently to achieve performance
gains. In case of OpenMP, the programmer is in charge of creating,
using and closing parallel regions (fork-join parallelism) and of
synchronising threads with MPI calls to implement a correct
communication pattern. The process of adding thread-level parallelism
to an MPI application is characterised by a sequence of ``looking for
code sections with significant duration'' and then ``adding
annotations to parallelise these code sections''. This is strenuous
and as a result, hybrid applications often end up with interleaved
execution \pmf{of concurrent computation phases (OpenMP) and
  communication phases with less parallelism (MPI), which limits the
  maximum speed-up of the entire application in accordance to Amdahl's
  law.  We call this two-phase programming.}

\pmf{It can be readily seen that it becomes necessary to define} a
methodology \pmf{that helps} developers to avoid the limitations of
two-phase programming, allowing a more natural coexistence between MPI
\pmf{and shared-memory programming models}.

\subsection{From processes to tasks}

In this work we present HDOT, \emph{hierarchical domain
  over-decomposition with tasking}, a \pmf{methodology} that
simplifies hybrid programming and improves the execution performance
of such applications.

HDOT applies the domain decomposition of an existing MPI application at process-level to thread-level. At \pmf{this} thread-level, domains are decomposed into subdomains and \pmf{executed} by tasks. HDOT maximises pattern and code reuse and adds the advantages of tasking. Tasking simplifies the development of hybrid applications by eliminating two-phase programming and by reducing the complexity of synchronisation of parallel work and MPI calls.

The HDOT methodology defines a top-down approach where the developer uses tasks to encapsulate work from a coarse-grained level down to tasks as small as individual send and receive MPI messages. This includes the definition of subdomains and support of common algorithmic patterns such as reductions and global variables.

\pmf{The OmpSs-2 programming model (see Section~\ref{subsec:ompss-2})} eliminates the notion of threads and emphasises the use of a task as a building primitive for concurrency.  Tasks are expressed declaratively and can encapsulate computation and MPI communication alike. The task-aware MPI (TAMPI) library (see Section~\ref{subsec:tampi}) ensures the correct execution of MPI calls within tasks. Finally, the programming model's support for data-flow programming allows a streamlined execution of computation and communication tasks.

In the next chapter we provide useful background information on today's challenges of MPI and hybrid programming, give a short introduction to OmpSs-2 and TAMPI, and discuss the related work. \rev{Next,} we present the aforementioned methodology and apply it on a set of applications. Finally, we present performance results achieved \rev{with MPI+OmpSs-2, MPI+OpenMP and MPI-only implementations.}

%% file: tex/background.tex
\section{Background}
\label{sec:background}

\subsection{MPI and OpenMP}
\label{subsec:threading_mpi}

Threading allows to reduce the number of \pmf{MPI} processes per node
while maintaining the same degree of concurrency of the application.
\pmf{OpenMP}~\cite{openmp} is a widely known multi-threaded
shared-memory programming model.  It allows to parallelise
applications by using a set of compiler directives, library routines
and environment variables.  OpenMP is flexible and supports most
parallel patterns: from \xt{the traditional work-sharing} approach
\pmf{to the OpenMP accelerator and tasking models}.  However, taking a
closer look at existing codes shows that the prevalent use of OpenMP
is the expression of data-parallel algorithms with an execution model
called fork-join.  Since OpenMP 3.0~\cite{openmp30}, \pmf{it is
  possible to express} dynamic task parallelism through task
generating constructs (including {\tt task}, {\tt taskgroup}, {\tt
  taskyield} and {\tt taskwait}). A task is a unit of work used to
express portions of code that could be executed concurrently by the
participant threads \rev{in accordance to certain restrictions.}

\begin{lstfloat}
\centering
\begin{lstlisting}[
   caption={Representative \pmf{MPI-only} application with interleaved point-to-point and collective communication.},
   label={background:code1}
  ]
int N = rank;               // assign process ID 
T D = getDomain(Domain, N); // set up domain for current process
for(auto t : timesteps){
  comm(D);
  f(D); g(D);
  collective_comm(D);
  h(D);} 
MPI_Barrier(...); // synchronise processes

\end{lstlisting}
\end{lstfloat}

Code~\ref{background:code1} shows an \pmf{MPI-only} application where
a sequence of functions is called in a simulation loop over a
predefined set of time steps. \pmf{In this code example, {\tt D} is an
  object that handles data associated with the current MPI domain
  whilst} {\tt comm} and {\tt collective\_comm} contain point-to-point
and collective MPI communication, respectively. Once the execution of
the simulation loop finishes, processes synchronise at an {\tt
  MPI\_Barrier}. Code~\ref{background:code2} shows the same
\pmf{algorithm} enhanced with OpenMP. As the number of processes is
reduced in favour of threading, the compute functions {\tt f}, {\tt g}
and {\tt h} are now placed inside a {\tt parallel} region and
executed on each thread (fork). Threaded execution requires to adjust
the implementations of these functions, hence we name them {\tt
  f\_parallel}, {\tt g\_parallel} and {\tt h\_parallel}. At the end of
each parallel region, an implicit barrier synchronises threads
(join). Such thread synchronisation is necessary in order to maintain
the correct ordering of communication and computation.

\begin{lstfloat}
\centering 
\begin{lstlisting}[ caption={A hybrid application implemented
    with MPI and OpenMP showing two parallel regions and the
    respective synchronisation points at thread- and process-level.},
  label={background:code2}, ]
int N = rank;               // assign process ID
T D = getDomain(Domain, N); // set up domain for current process
for(auto t : timesteps){
  comm(D);
  #pragma omp parallel
  {f_parallel(D); g_parallel(D);} // synchronise threads (join)
  collective_comm(D);
  #pragma omp parallel
  h_parallel(D);}                 // synchronise threads (join)
MPI_Barrier(...);                 // synchronise processes
\end{lstlisting}
\end{lstfloat}

\begin{figure}[tp]
  \begin{minipage}[c]{\textwidth}
    \begin{center}
      \hspace*{\fill}
      \subfloat[]{
        \includegraphics[height=4cm]{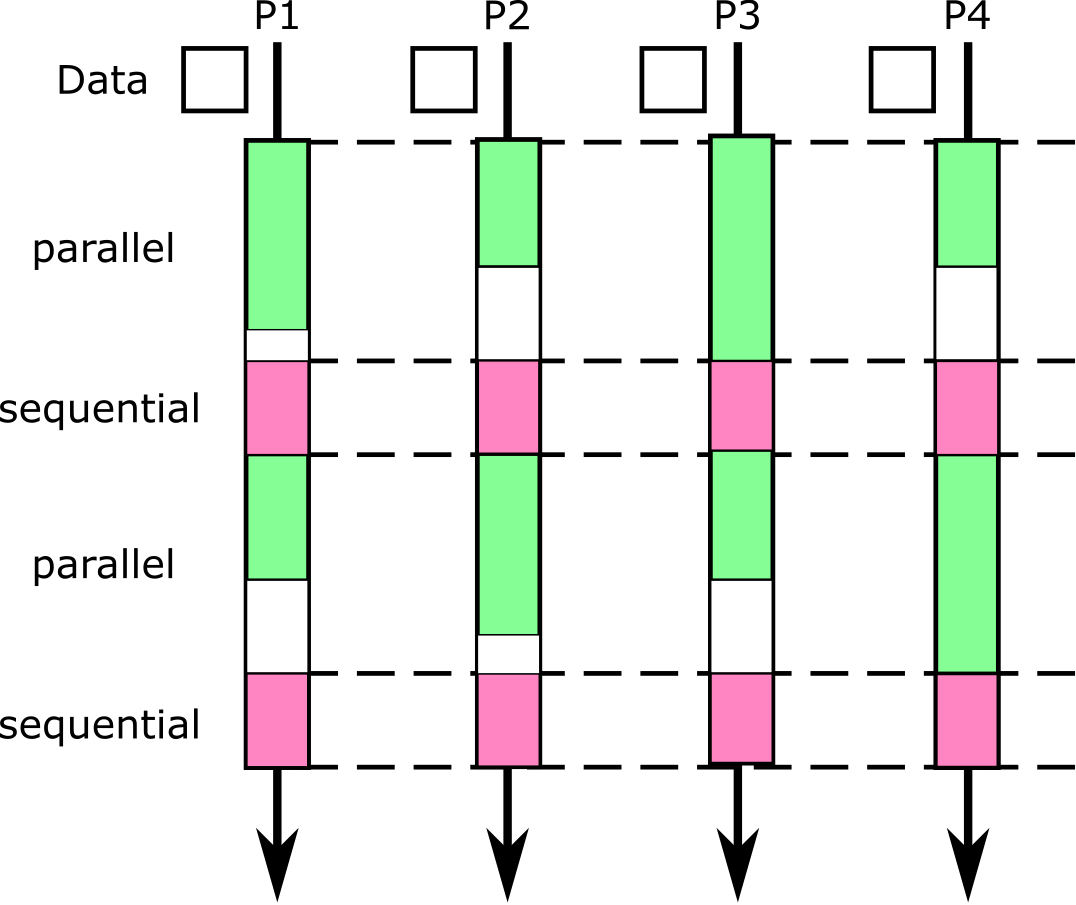}
      }
      \hfill
      \subfloat[]{
        \includegraphics[height=4cm]{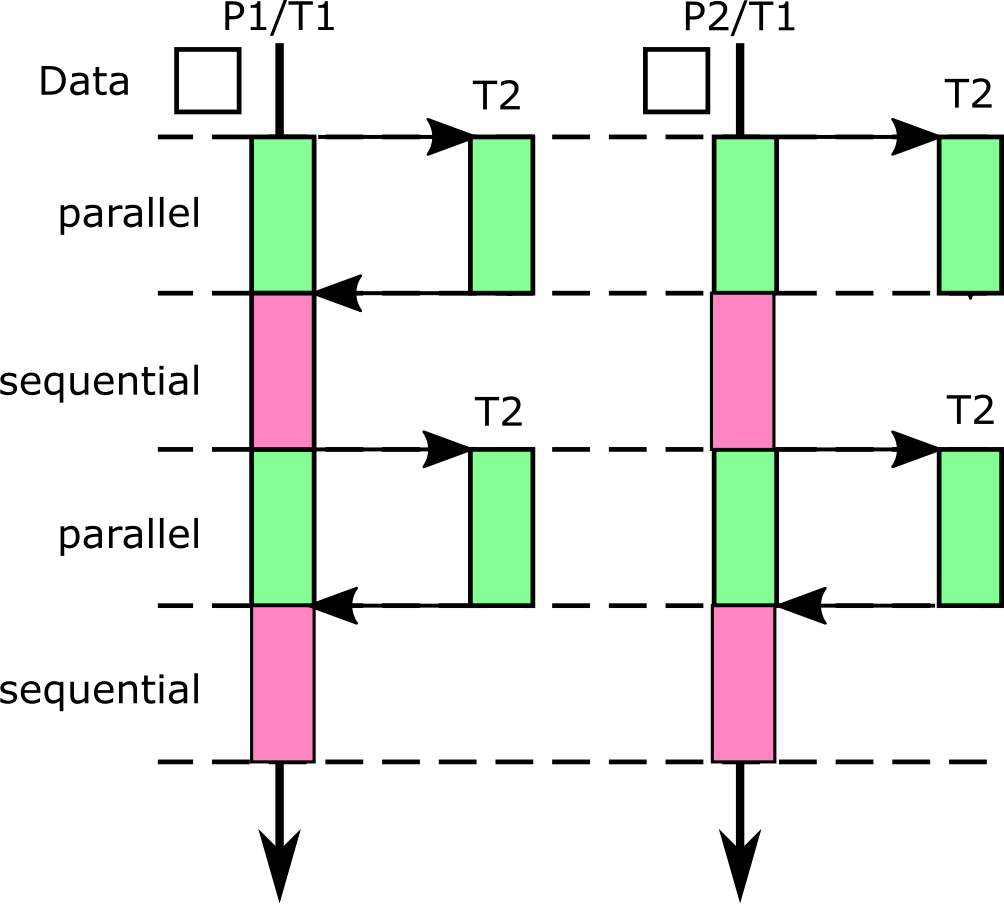}
      }
      \hspace*{\fill}
    \end{center}
  \end{minipage}
  \caption{Execution diagram of an application with interleaved
    communication and computation \pmf{using (a) MPI-only and (b)
      MPI+OpenMP implementations}. The hybrid version illustrates the
    fork-join parallel execution of the compute functions (parallel)
    and its ability to balance load among threads.}
  \label{background:pic1}
\end{figure}

Figure~\ref{background:pic1} shows an execution diagram \rev{of} the
fork-join parallelism of \rev{MPI-only and MPI+OpenMP
  implementations}. The execution diagram also illustrates, \rev{on
  the one hand,} the additional amount of data copies \rev{of the
  MPI-only version} (multiple data objects) \rev{and, on the other
  hand,} the ability of load-balancing of the hybrid version.

\subsection{Tasking with OmpSs-2}
\label{subsec:ompss-2}

OmpSs-2~\cite{ompss-2} (OMP SuperScalar v2) is a high-level,
task-based, parallel programming model for shared-memory systems
developed at the Barcelona Supercomputing Center (BSC). It consists of
a language specification, a source-to-source compiler for C, C++ and
Fortran as well as a runtime.  OmpSs-2 defines a set of directives
\rev{for} a descriptive expression of tasks.  Further, it allows the
programmer to annotate task parameters with {\tt in}, {\tt out} and
{\tt inout} clauses that correspond to the input, output or the
combination of input-output access type semantic of these parameters
within a task. Each access type clause receives a list of parameters:

\begin{minipage}{\textwidth}
\begin{lstlisting}[style=customFortran2]
#pragma oss task [in(par-list) | out(par-list) | inout(par-list)]
\end{lstlisting}
\end{minipage}

The access type establishes a producer-consumer relationship between tasks, also called task dependency or data-flow. With this information the runtime is capable of scheduling tasks automatically that maintain correctness of code while alleviating the programmer of implementing manual synchronisation. Furthermore, the {\tt taskwait} construct allows task synchronisation and instructs the calling thread to wait on all previously created tasks. While this is similar to tasking in the recent specification of OpenMP, the OmpSs-2 runtime implements a different execution model.

In OmpSs-2, every application starts with a predefined set of execution resources, \rev{that is,} an explicit {\tt parallel} region does not exist. This view avoids the exposure of threading to the programmer as well as the requirement to handle an additional scope \rev{as in OpenMP}. At compile time, the OmpSs-2 compiler processes pragma annotations and generates an intermediate code file. This file includes \pmf{both user code and additional code} for task generation, synchronisation and error handling. In the final step of compilation, OmpSs-2 invokes the native compiler to create a binary file.

At runtime, the {\tt main} function is executed, which creates tasks and stops at (explicit or implicit) synchronisation points. Task creation is composed of two parts: (i) the creation of the task object itself that carries all its descriptive information and (ii) its data dependencies. Once a task object has been created, the runtime inspects the dependency graph to determine the relationship with respect to previously created tasks. If a dependency has been found, a representative node is added to the graph. In the opposite case, the task is placed into a ready-queue. Tasks in the ready-queue are picked up by worker threads, removed from the queue and executed.

Interestingly, one question arises: is it possible to leverage the
dynamic, data-flow driven task execution in OmpSs-2 to improve the
performance of MPI-only applications? In the following section we
present TAMPI, a task-aware MPI library, which enables a natural
coexistence of the OmpSs-2 shared-memory programming model with the
MPI \pmf{distributed-memory specification}.

\subsection{Task-aware MPI}
\label{subsec:tampi}

The task-aware MPI (TAMPI)
library~\cite{Sala2018,KSala2018} extends the
functionality of standard MPI libraries by providing new mechanisms
for improving the interoperability between parallel task-based
programming models, such as OpenMP or OmpSs-2, and both blocking and
non-blocking MPI operations.

By following the MPI specification, programmers must pay close
attention to avoid deadlocks that may occur in hybrid applications
(e.g.\ MPI+OpenMP) where MPI calls take place inside tasks. This is
given by the out-of-order execution of tasks that consequently alter
the execution order of the enclosed MPI calls.  The TAMPI library
ensures a deadlock-free execution of such hybrid applications by
implementing a cooperation mechanism between the MPI library and the
parallel task-based runtime system.

TAMPI provides blocking~\cite{Sala2018} and
non-blocking~\cite{KSala2018} modes. The blocking mode targets the
efficient and safe execution of blocking MPI operations
(e.g.\ MPI\_Recv) from inside tasks, while the non-blocking mode
focuses on the efficient execution of non-blocking or immediate MPI
operations (e.g.\ MPI\_Irecv), also from inside tasks.

TAMPI is compatible with mainstream MPI implementations that support the
MPI\_THREAD\_MULTIPLE threading level, which is the minimum requirement to
provide its task-aware features. This library is a refinement of the hybrid
MPI+SMPSs approach presented in~\cite{smpss_mpi}.  We consider it as a key
feature towards achieving performance and scalability of hybrid applications on
modern systems today.  It is worth mentioning that it is still the user's
responsibility to prevent races when threads within the same application post
conflicting communication calls. Furthermore, we would like to emphasise that
the development of hybrid applications based on tasks can be fully achieved
without using this library, but at the expense of a greater programming effort.
This is briefly discussed in Section~\ref{sec:creams}.

\subsection{Related work}
\label{subsec:related}

\xt{ MPI-based parallelisation approaches try to improve performance
  and scalability of applications by overlapping communication and
  computation~\cite{Doerfler2006, Hoefler2007, Graham2010} and by
  accelerating the execution of the critical path of the
  program~\cite{Yang1988, Hollingsworth1998, Schulz2005, Schmitt2014}.

Although the MPI-only approach is still one \rev{of} the most popular
among the set of parallelisation alternatives within the high
performance computing community, programmers will have to exploit
hybrid solutions (MPI+X) that allow a better use of hardware resources
in order to reach the so-called exascale era~\cite{Thakur_mpiat}.  In
general, these hybrid approaches also try to enhance the application
behaviour by overlap and critical path programming
techniques~\cite{Chatterjee2013,smpss_mpi}.

SMPSs~\cite{smpss_mpi} enables communication and computation overlap
by taskifying MPI calls and adding a restart functionality which
allows re-scheduling tasks waiting for a certain condition (similarly
to a mandatory {\tt taskyield} with extra scheduler semantics). In our
approach, this functionality is transparently provided by the TAMPI
library. Furthermore, in SMPSs the programmer needs to explicitly
split a blocking call into a non-blocking call to issue the
communication request and wait for the data, while TAMPI does not
require this transformation from the programmer.

HCMPI~\cite{Chatterjee2013} unifies the Habanero-C intranode task
parallelism with MPI internode parallelism and extends the tasking
model by allowing regular computation tasks to create asynchronous
communication tasks. In HCMPI's data-flow model, synchronisation
between computation and communication tasks can be achieved through
the specific {\tt await} clause and other specific MPI-like services
(e.g.\ {\tt wait}, {\tt waitall} or {\tt waitany}).

Both SMPSs and HCMPI require a communication thread dedicated to
execute MPI calls, while our methodology does not force how to
implement \rev{the} communication progress. \rev{Moreover}, both
approaches are focused exclusively on the communication pattern,
independently \rev{of} how the computational decomposition \rev{is}
carried out. We consider \rev{that} these two elements (computation
and communication) are so dependent the one to \rev{the} other that
they should be interrelated when applying any parallel decomposition
approach.

The methodology presented in this work is also aligned with those
build on top of task-based runtime systems (also referred to as
user-level threads) as a second level of parallelism.  The main reason
for using this approach is that it indeed represents a more natural
way to convert MPI code to tasking of a shared-memory parallel
programming model. Among those related works proposing
task-parallelism, we would like to highlight the task parallel
over-decomposition (TPOD) approach~\cite{Barrett2015} due its
similarity \rev{to} the present work.

TPOD combines computation and communication operations in the same
task in a way that maintains the traditional MPI coding style. This
strategy is based on over-decomposing the domain that has been
assigned to a set of cores (i.e.\ MPI processes) into a set of smaller
subdomains (i.e.\ tasks). The main difference of TPOD tasks with
respect to regular tasks \rev{is found} when the code reaches a
blocking communication call. In \rev{this situation} the task may be
swapped out, allowing another task to be swapped in and continue with
the execution. At some point, the communication-blocked task \rev{will
  be} swapped back in and \rev{continue with its normal execution}.

Although the TPOD approach also uses tasks to overlap computation and
communication phases, it relies on global barriers between different
iterations. This strategy restricts the degree of potential
concurrency that the application can reach, being better to rely on
fine-grained synchronisation such as the task dependency
system. MPI+OmpSs-2 naturally exploits nesting capabilities to balance
the trade-off between grain size and the degree of concurrency based
on a top-down approach.  }

%% file: tex/methodology.tex
\section{HDOT: Hierarchical Domain Over-decomposition with Tasking}
\label{sec:methodology}

HDOT leverages the parallelisation and domain decomposition schemes of the original MPI application by promoting their reuse on task-level. In this scheme, domains are split hierarchically, first at process-level \pmf{(MPI)}, down to task-level \pmf{(shared memory)} at which domains \pmf{are referred to as} subdomains.

Applying a hierarchical over-decomposition with tasking follows a set of underlying ideas. Firstly, it minimises requirements for code changes and allows reuse of original MPI code and its data partitioning. Secondly, it decouples tight synchronisation between units of work and MPI communication.  And lastly, it emphasises the use of a top-down approach where concurrency is added on different nesting levels to create opportunity for concurrency at a small development effort. We discuss this in three steps as follows.

\subsection{Taskifying code}

Let us continue with Code~\ref{background:code2}. In this code example, the maximal speed-up is limited by the fork-join pattern with sequential sections and the tight synchronisation between computation and communication. To apply the \pmf{HDOT} methodology, we start with the original\pmf{, MPI-only Code~\ref{background:code1}}. In the first step, we taskify code sections. That is, we start adding the OmpSs-2 task pragmas to the code, stating that the enclosed code is adept for concurrent execution. \pmf{It is important to point out that, by taskifying most of the application code, the number of synchronisation points gets reduced}. The execution of tasks in the correct order is guaranteed by following the application data-flow. Code~\ref{methodology:code1} is based on the previous example but now includes tasks with their respective data dependencies. 
\begin{lstfloat}
\centering
\begin{lstlisting}[
  caption={Sample code showing the top-down parallelisation of computation and communication of an MPI application with OmpSs-2 tasks.},
  label={methodology:code1},
  ]
int N = rank;             // assign process ID 
T D = getDomain(data, N); // set up domain for current process
for(auto && t : timesteps){
  #pragma oss task inout(D) 
  comm(D);
  #pragma oss task inout(D) 
  {f(D); g(D);}
  #pragma oss task inout(D) 
  collective_comm(D);
  #pragma oss task inout(D) 
  h(D);} 
#pragma oss taskwait // synchronise threads
MPI_Barrier(...);    // synchronise processes

\end{lstlisting}
\end{lstfloat}

In this case, the \pmf{non-blocking} execution of MPI calls by
including them in tasks and in the application data-flow is an
important feature towards scalability and programmability.  \pmf{In
  Code~{\ref{methodology:code1}},} dependencies of the type {\tt
  inout} serialise the execution of the three tasks. The code thus
requires a finer task granularity \pmf{via subdomains}, which is
discussed in the following section.

\subsection{Adding subdomains}

To increase the degree of concurrency of the application shown in
\pmf{Code}~\ref{methodology:code1}, we implement what we refer to as a
domain over-decomposition that splits domains into subdomains of
arbitrary sizes. \pmf{On the one hand, a} domain is a
\rev{\emph{physical}} data partition implemented originally by an
\pmf{MPI} application\pmf{; on the other hand, a} subdomain is a
\rev{\emph{virtual}} data partition implemented for the use of a
shared-memory programming model with the aim of increasing the
node-level parallelism.  Subdomains follow the same idea of data
partitioning found on process level. In case of subdomains, tasks and
task functions operate on smaller, process-local data with occasional
communication. The implementation of those functions remains
unchanged.

\begin{lstfloat}
\centering
\begin{lstlisting}[
  caption={\pmf{Subdomains allow to reuse the domain partitioning at process-level but also require to check for boundary subdomains and the use of weak dependencies.}},
  label={methodology:code2},
]
int N = rank;             // assign process ID
T D = getDomain(data, N); // set up domain for current process
for(auto && t : timesteps){
  #pragma oss task weakinout(D)
  for(auto && subdomain : D.getSubDomains()){ // subdomain loop
    if(subdomain.isBoundary()){
      #pragma oss task inout(subdomain)
      comm(subdomain);}}
  #pragma oss task weakinout(D)
  for(auto && subdomain : D.getSubDomains()){ // subdomain loop
    #pragma oss task inout(subdomain)
    {f(subdomain); g(subdomain);}}
  #pragma oss task inout (D[0:D.getNumSubDomains()-1])
  collective_comm(D[0;NB-1]);
  #pragma oss task weakinout(D)
  for(auto && subdomain : D.getSubDomains()){ // subdomain loop
    if(subdomain.isBoundary()){
      #pragma oss task inout(subdomain)
      h(subdomain);}}}
#pragma oss taskwait // synchronise threads
MPI_Barrier(...);    // synchronise processes
\end{lstlisting}
\end{lstfloat}

Code~\ref{methodology:code2} shows the code changes made in order to accommodate subdomains. It can be seen that we have added an additional task nesting level\pmf{, i.e.\ subdomain loops}. On the inner nesting level, new tasks are created in for-loops where the number of inner tasks corresponds to the number of subdomains. To add these tasks to the data-flow of the application, each newly created task defines an {\tt inout} dependency over an individual subdomain. However, since encapsulating tasks from the original code would still serialise the execution between loops, we apply the concept of \pmf{weak dependencies via the prefix {\tt weak}}.

Weak dependencies~\cite{nanos6} is a key feature of the OmpSs-2 programming model to allow a top-down parallelisation of code. \pmf{They break} coarse-grained dependencies between tasks under the assumption that inner tasks will fulfill the dependency requirements. \pmf{This results} in fine-grained dependencies between tasks where a communication task over a subdomain can be immediately \pmf{executed} once the prior computation task over that subdomain has finished.  Towards the end of the code sample, a collective MPI operation accesses all subdomains and therefore defines an {\tt inout} dependency over all of them.  \pmf{Furthermore,} since not all subdomains are equal in their correspondence to the geometric position in the original domain, we have added the \pmf{condition {\tt isBoundary} to check whether the subdomain type requires to communicate halo data with MPI}.

\pmf{Finally, it is worth noting two important aspects about subdomains
  and their associated task granularity.  Firstly, on task-based
  runtime systems the number of tasks (which is inversely proportional
  to the task granularity) should be preferably larger than the number
  of cores (even an order of magnitude larger) available per MPI
  process to guarantee good scalability properties, that is, to
  guarantee enough parallelism without incurring into runtime scheduling
  costs.  On the other hand, network saturation could become an issue
  with a fine granularity, for instance, when tasks communicate
  message sizes smaller than 1~KB, approximately, on an InfiniBand
  network.  In such case, the task granularity associated with
  communication must be increased whilst maintaining the same
  granularity for computation tasks.}

\subsection{Including support for common programming patterns}

In many scientific codes, developers use reductions and global
(static) variables. \pmf{Reductions are operations that possess an
  identity element and can be processed in parallel since} each
concurrent unit of work can initialise a private set of operands to
the neutral element \rev{following the same approach adopted by
  OpenMP}. We make use of these properties in HDOT and show how
reductions are computed on task- and process-level. In those cases the
reduction value represents either an intermediate result at task-level
or a final result on process-level.

\begin{lstfloat}
\centering 
\begin{lstlisting}[ caption={Support for reductions and
    static variables requires the addition of the reduction clause and
    local variables, respectively.}, label={methodology:code3},
]
double residual = DOUBLE_MAX; // assign value to reduction variable
while(residual > norm){
  double rlocal = DOUBLE_MAX;
  #pragma oss task weakinout(D) inout(rlocal)
  for(auto && subdomain : D.getSubDomains()){ // subdomain loop
    #pragma oss task inout(subdomain) reduction(MAX:rlocal)
    rlocal = MAX(rlocal, f(subdomain));}
  #pragma oss task in(rlocal) inout(residual)
  MPI_Allreduce(rlocal, residual ...);}       // MPI collective comm.
#pragma oss taskwait // synchronise threads
MPI_Barrier(...);    // synchronise processes
\end{lstlisting}
\end{lstfloat}

Code~\ref{methodology:code3} is a continuation of the code samples previously shown; however, we have removed irrelevant code lines.  \pmf{The {\tt reduction} clause} instructs the runtime system to provide a thread-safe storage such that the execution of any two concurrent tasks is data-race free. Once all participating tasks complete, the task \pmf{calling the {\tt MPI\_Allreduce} function} can be executed. In OmpSs-2, a reduction creates an input-output dependency implicitly.

In order to maintain the dependency between computation and communication tasks, we have added a stack-local reduction variable called {\tt rlocal}. This creates a dependency between the computation tasks that perform the reduction and the communication task that perform the {\tt MPI\_Allreduce} which defines an input over that variable.

Algorithms often use global variables to store the state of the
simulation or geometric properties describing a domain. However, an
over-decomposition of the simulation domain requires stack-local
variables that hold pointers to such data structure for each task. For
this purpose, Codes~\ref{methodology:code2}--\ref{methodology:code3}
implement \pmf{the functions} {\tt getSubDomains} and {\tt
  getNumSubDomains} that return a set of subdomains and their number,
respectively. Using pragma annotations similar to {\tt firstprivate}
for custom data types is not optimal as this typically invokes the
copy constructor of that object resulting in replication of
potentially large data.

%% file: tex/heat.tex
\section{Applications}

\pmf{In this section we present the performance results obtained for
  two benchmarks, Heat2D and HPCCG, and one large application, CREAMS.
  The results were obtained on the MareNostrum 4 supercomputer located
  at the Barcelona Supercomputing Center (BSC).  Each node of
  MareNostrum 4 is equipped with a dual-socket Intel Platinum 8160 CPU
  featuring 24 cores per processor, 96 GB of main memory and Intel's
  Omni-Path HFI interconnect network.}

\subsection{Heat2D benchmark}
\label{sec:heat}

Heat2D is a popular \pmf{benchmark} that simulates heat diffusion in two dimensions.  At each time step, a blocked Gauss-Seidel iterative solver approximates a solution of the Poisson equation. The continuous problem is discretised with finite differences. The resulting calculation for each discrete point is defined as: \pmf{$U_{i,j} = (U_{i+1,j} + U_{i-1,j} + U_{i,j+1} + U_{i,j-1})/4$}.

Heat2D is characterised by its stencil operation where each \pmf{point update} requires the access to the corresponding left, right, top and bottom \pmf{neighbouring points}. 

\begin{lstfloat}
\begin{lstlisting}[
    caption={\pmf{Pseudocode of the Heat2D benchmark.}},
    language=C,
    label={code:heat-MPI}
  ]
N = rank;       // assign process ID
M = ranks;      // assign total number of processes
T D = setup(N); // setup simulation domain for current process
T first, last;  // halos
while(residual > norm){
  if(N > 0) {send(first, N-1); receive(last, N-1);}
  if(N < last) {receive(first, N+1);}
  solveBlock(D, first, last, residual ...);
  if(N < last) {send(last, N+1);}
  allReduce(residual ...);}
\end{lstlisting}
\end{lstfloat}

\begin{figure}[tp]
  \begin{minipage}[c]{0.985\textwidth}
    \begin{minipage}[c]{0.27\textwidth}
      \begin{center}
        \subfloat[]{
          \includegraphics[width=\textwidth]{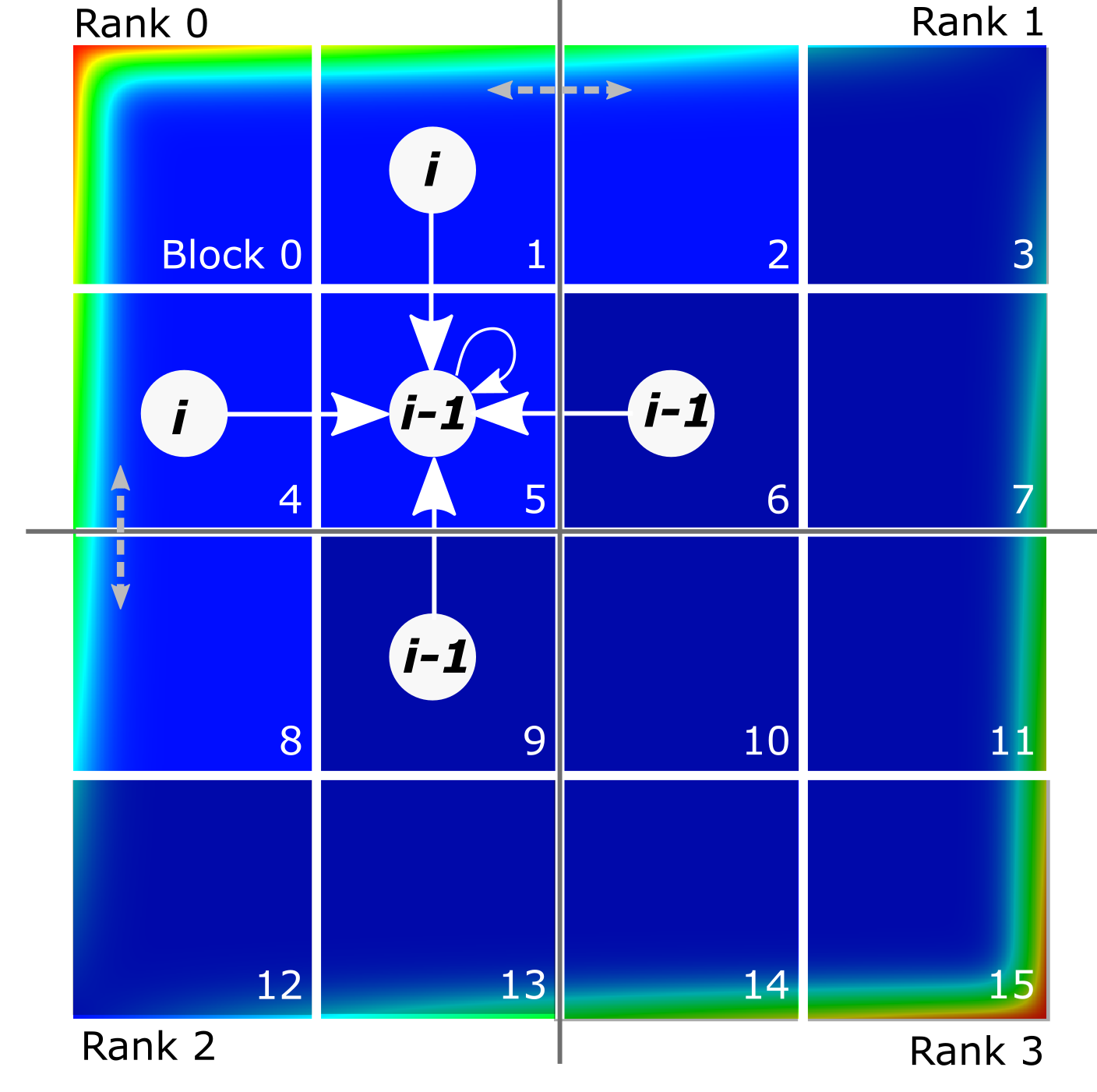}\label{fig:heat-pic1}
        }
      \end{center}
    \end{minipage}
    \hfill
    \begin{minipage}[c]{0.70\textwidth}
      \begin{center}
        \subfloat[]{
          \includegraphics[width=\textwidth]{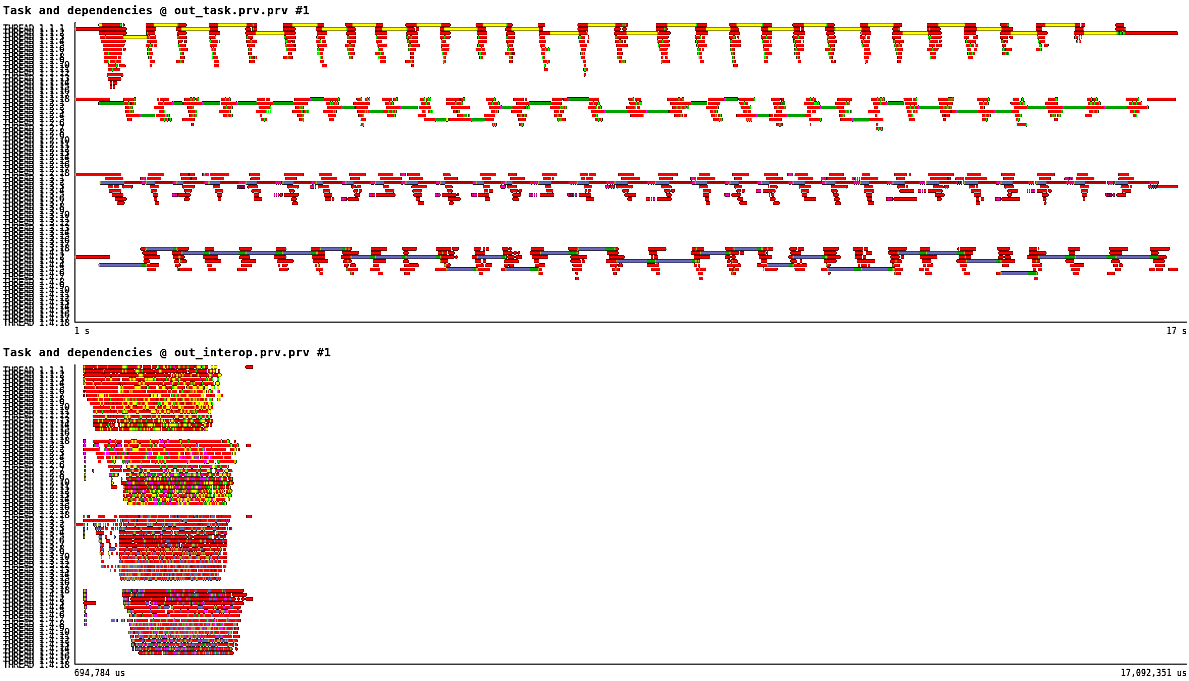}\label{fig:heat-pic3}
        }
      \end{center}
    \end{minipage}
  \end{minipage}
  \caption{\rev{(a) Heat2D stencil operation results in a wave-front
      progression on process- and thread-level (its efficient
      execution requires fine-grained dependency resolution as well as
      computation and communication overlapping); (b) Heat2D traces on
      four nodes of MareNostrum 4 show massive differences in resource
      utilisation and performance between a traditional fork-join
      approach using MPI+OpenMP (top) and our HDOT implementation
      using MPI+OmpSs-2 (bottom).}}
\end{figure}

\pmf{Code~\ref{code:heat-MPI} gives a basic idea of how the heat
  benchmark is implemented, including the communication of halos and
  the actual computation carried out by the function {\tt solveBlock},
  whilst} Figure~\ref{fig:heat-pic1} shows the stencil operation and a
visual representation of the progression of the algorithm mapped over
the same data set used for benchmarking later in this section.  In
this figure, the simulation space is divided in 4 MPI \rev{domains
  (ranks)}. During execution, each MPI rank \pmf{contains} four
subdomains (blocks) and each subdomain is updated by one \pmf{OmpSs-2}
task. Dependencies that originate from the stencil operation for each
element can be equally applied to define dependencies at block or
process level. That is, \pmf{MPI} processes 1 and 2 can only start
when blocks 1 and 4 of process 0 have finished computation and when
the associated communication tasks (marked by the dotted double-sided
arrows in Figure~\ref{fig:heat-pic1}) have completed as well.
\pmf{Hybrid code implementations of this particular benchmark (not
  shown here for the sake of conciseness) can be found in an early
  form in~\cite{Sala2018}.}

\pmf{Figure~\ref{fig:heat-pic3} shows a trace of the execution of the
  Heat2D application on \rev{four} nodes of the MareNostrum 4
  supercomputer implemented with MPI+\rev{OpenMP} (top) and
  MPI+\rev{OmpSs-2} (bottom).  It can be readily seen that the
  implementation using OpenMP (fork-join execution model) is not capable of generating enough
  parallelism in order to \pmf{maintain all the processor cores
    busy}.}

\begin{table}
 \footnotesize
 \centering
 \begin{tabular}{c S[table-format=5.0] c S[table-format=5.0] S[table-format=2.1] cc}
  \toprule
  Nodes & $\mathrm{Perf._{M+OSs2}}$ & $\mathrm{Perf._{M+OMP}}$ & $\mathrm{Perf._{M}}$ & $\mathrm{S._{M+OSs2}}$ & $\mathrm{S._{M+OMP}}$ & $\mathrm{S._{M}}$ \\
  \hline
  \phantom{0}1 & 6561   & 2836 & 2479  & 2.6  & 1.1 & 1.0 \\
  \phantom{0}2 & 12729  & 2096 & 4306  & 5.1  & 0.8 & 1.7 \\
  \phantom{0}4 & 25231  & 2134 & 6809  & 10.2 & 0.9 & 2.7 \\
  \phantom{0}8 & 48302  & 2757 & 9555  & 19.5 & 1.1 & 3.9 \\
            16 & 99032  & 4259 & 11956 & 39.9 & 1.7 & 4.8 \\
            32 & 145547 & 5839 & 13211 & 58.7 & 2.4 & 5.3 \\
  \bottomrule
\end{tabular}
 \caption{\rev{Heat2D execution performance measured in giga updates
     per second (more is better) for MPI+OmpSs-2 (M+OSs2), MPI+OpenMP
     (M+OMP) and MPI-only (M) implementations.  Speed-up numbers (S)
     are normalised by the execution performance of the MPI-only
     version on one node.  For the MPI-only version the number of MPI
     processes or ranks equals the total number of cores.}}
\label{heat:table1}
\end{table}

\pmf{Table~\ref{heat:table1} show\rev{s} the performance results for
  three parallel implementations of this application using up to 32
  nodes.  The implementation with OmpSs-2 and the TAMPI library allows
  to streamline the interleaved execution of computation and
  communication tasks \pmf{shown in Code~\ref{code:heat-MPI} and
    avoid} any form of thread-local (OpenMP) or process-wide (MPI)
  synchronisation.  On the other hand, the hybrid MPI+OpenMP
  implementation requires the presence \rev{of} a thread barrier ({\tt
    \#pragma omp barrier}) at the end of each loop that processes the
  local blocks ({\tt solveBlock}). This way, any communication task
  for a particular halo can only be executed once computations of all
  blocks finish their execution.  Similarly, the MPI-only
  implementation shows significant slow-downs due to the use of
  \pmf{blocking} MPI calls in order to maintain a correct ordering of
  computation and halo exchange}.  \xt{Authors' previous
  publications~\cite{Sala2018,KSala2018} present a deeper study about
  the performance of each of these versions, including the results
  obtained by directly using the non-blocking MPI services.}




%% file: tex/HPCCG.tex
\subsection{HPCCG benchmark}
The high performance computing conjugate gradient
(HPCCG)\cite{heroux2009improving} benchmark \pmf{consists of a}
synthetic sparse linear system that is mathematically similar to
finite element, finite volume or finite difference discretisations of
a three-dimensional heat diffusion problem on a semi-regular grid. The
problem is solved using domain decomposition with an additive Schwarz
preconditioned conjugate gradient method where each subdomain is
preconditioned using a symmetric Gauss-Seidel sweep.

The HPCCG benchmark generates a three-dimensional partial differential
equation model problem and computes \rev{the} preconditioned conjugate
gradient iterations \rev{from} the resulting sparse linear system. The
global domain dimensions are $n_x \times n_y \times (n_z \times n_p)$,
where $n_x \times n_y \times n_z$ are the local subgrid dimensions
assigned to each MPI process and $n_p$ the number of MPI processes.
\pmf{In this work,} each MPI process or domain is stacked in the
$z$-direction.

\begin{lstfloat}
\begin{lstlisting}[
    caption={HPCCG body loop using an MPI-only implementation.},
    language=C,
    label={code:HPCCG_loop}
  ]
double rtrans, oldrtrans, alpha, beta;
double r[], p[], Ap[], x[];
HPC_Sparse_Matrix A;
for(i = 0; i < max_iter /* and normr > tolerance */; ++i){
  if(i == 1){waxpby(1.0, r, 0.0, r, p);} // p = 0.0*r + 1.0*r
  else{
    oldrtrans = rtrans;
    ddot(r, r, &rtrans);          // dot_product(r, r) and MPI_Allreduce(rtrans)
    beta = rtrans/oldrtrans;
    waxpby(1.0, r, beta, p, p);}  // p = 1.0*r + beta*p
  normr = sqrt(rtrans);           // early exit
  exchange_externals(A, p);       // communication
  HPC_sparsemv(A, p, Ap);         // matrix_x_vector(A, p) = Ap
  ddot(p, Ap, &alpha);            // dot_product(p, Ap) and MPI_Allreduce(alpha)
  alpha = rtrans/alpha;
  waxpby(1.0, x, alpha, p, x);    // x = 1.0*x + alpha*r
  waxpby(1.0, r, -alpha, Ap, r);} // p = 1.0*r - alpha*r
\end{lstlisting}
\end{lstfloat}

Halos are computed at the beginning of the program and exchanged at each iteration of the algorithm, as shown in line 14 of Code~\ref{code:HPCCG_loop}.  On the one hand, the MPI+OpenMP approach consists of parallelising each operation using the {\ttfamily pragma omp parallel for} clause. On the other hand, the MPI+OmpSs-2 approach divides all the data structures in subdomains that can be executed in parallel via tasks.

\begin{lstfloat}
\begin{lstlisting}[
    caption={HPCCG (simplified) body loop using an MPI+OmpSs-2 implementation.},
    language=C,
    label={code:HPCCG_loop_ompss}
  ]
for(i = 0; i < max_iter /* and normr > tolerance */; ++i){
  if(i == 1){
    for(auto &&subd : D.getSubDomains()){
      #pragma oss task in(r[subd.from:subd.to]) out(p[subd.from:subd.to])
      waxpby(subd, 1.0, r, 0.0, r, p);}}
  else{
    for(auto &&subd : D.getSubDomains()){
      #pragma oss task in(r[subd.from:subd.to]) reduction(+:rtrans_L)
      ddot(subd, r, r, &rtrans_L);}
    #pragma oss task in(rtrans_L) out(rtrans, beta)
    {MPI_Allreduce(&rtrans_L, &rtrans); beta = rtrans/oldrtrans;}}
  exchange_externals(A, p); // tasks made inside
  for(auto &&subd : D.getSubSubDomains()){
    #pragma oss task in(p[subd.S:subd.E]) out(Ap[subd.S:subd.E]) reduction(+:alpha_L)
    {HPC_sparsemv(A, subd, p, Ap); ddot(subd, p, Ap, &alpha_L);}}
  #pragma oss task in(alpha_L) out(alpha)
  {MPI_Allreduce(&alpha_L, &alpha); alpha = rtrans/alpha;}}
\end{lstlisting}
\end{lstfloat}

A snipet of the hybrid \pmf{implementation of HPCCG} using HDOT is
illustrated in Code~\ref{code:HPCCG_loop_ompss}. It is worth noting
that some variables\pmf{, such as {\ttfamily rtrans} and {\ttfamily
    alpha},} need to be reduced in order to share them with other MPI
processes. Moreover, it is necessary to \rp{take out the} \pmf{call
  to} {\ttfamily MPI\_Allreduce} inside the function {\ttfamily ddot}
due to the need of a local reduction. The most computationally
demanding operation, i.e.\ {\ttfamily HPC\_sparsemv}, has been further
divided by using nesting within subdomains, allowing to expose more
parallelism.

\begin{figure}
\centering
\includegraphics[width=0.75\textwidth]{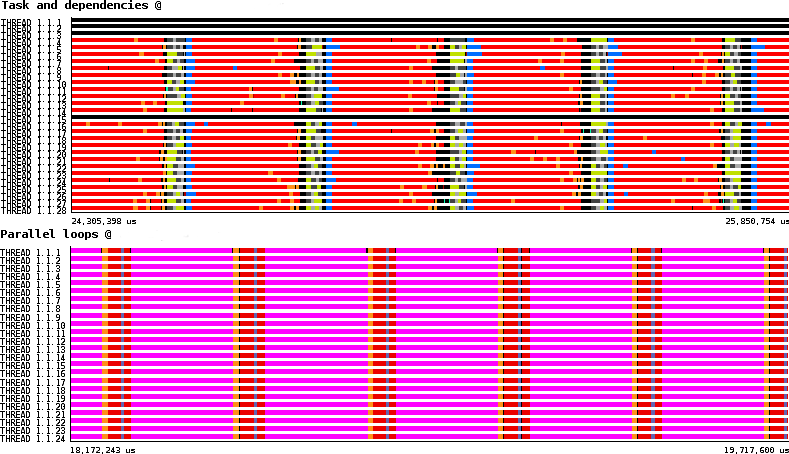}
\caption{\rev{HPCCG paraver traces for} MPI+OpenMP (top) \rev{and}
  MPI+OmpSs-2 (bottom) \rev{implementations obtained with} one node
  of MareNostrum 4.}
\label{fig:omp_vs_ompss}
\end{figure}

Figure~\ref{fig:omp_vs_ompss} shows a trace of \pmf{HPCCG implemented
  with MPI+OpenMP following a two-phase approach (top) and MPI+OmpSs-2
  following the HDOT approach (bottom)}.  The MPI+OmpSs-2 version has
a total of 32 subdomains, where nesting is applied on the {\ttfamily
  HPC\_sparsemv} operation in order to feed the 48 CPUs available in
one node.

\begin{figure}[tp]
  \begin{minipage}[c]{\textwidth}
    \begin{center}
      \hspace*{\fill}
      \subfloat[]{
        \includegraphics[height=4.35cm]{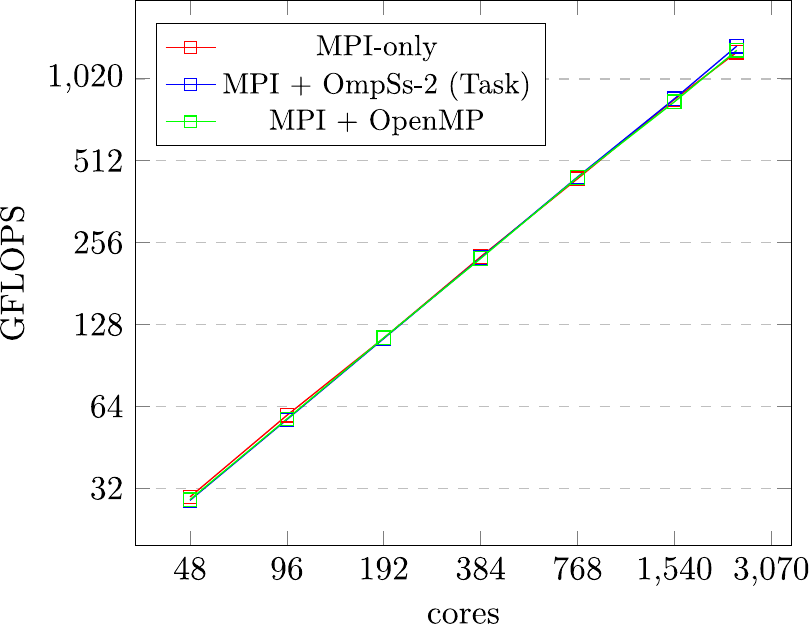}\label{fig:HPCCG-perf}
      }
      \hfill
      \subfloat[]{
        \includegraphics[height=4.35cm]{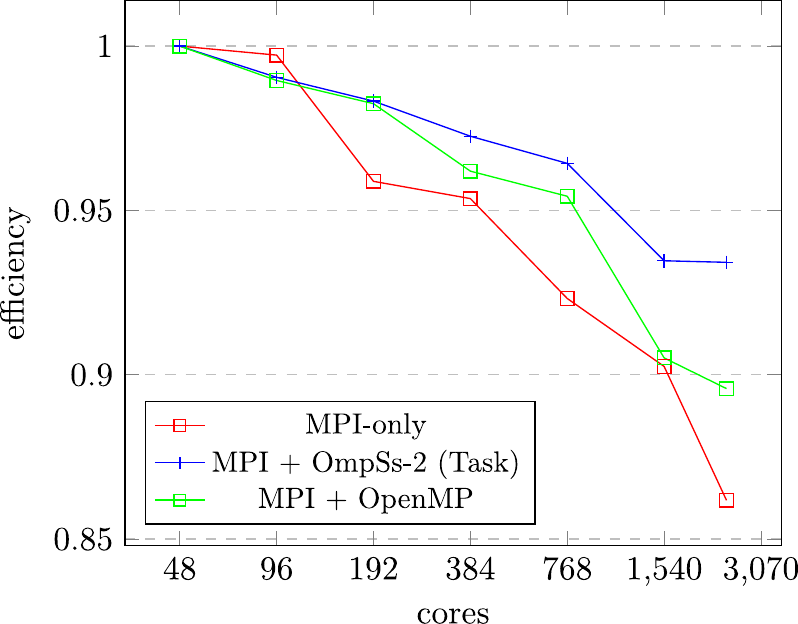}\label{fig:HPCCG-eff}
      }
      \hspace*{\fill}
    \end{center}
  \end{minipage}
  \caption{\rev{(a) HPCCG performace and (b) efficiency for different
      parallel implementations using up to \rev{50 nodes} (2400 cores)
      of MareNostrum 4.}}
\end{figure}

\rp {Figures~\ref{fig:HPCCG-perf}--\ref{fig:HPCCG-eff} show the
  performance and efficiency \rev{of HPCCG for} three parallel
  implementations running \rev{on} up to 50 nodes \rev{of MareNostrum
    4}.  The MPI+OmpSs-2 version allows \rev{interleaving} computation
  between subdomains by using the TAMPI library thus avoiding
  unnecessary waits, whilst the OpenMP version must wait for each
  parallel computation due to an implicit fork-join barrier.}




%% file: tex/creams.tex
\subsection{CREAMS application}\label{sec:creams}

The compressible reactive multi-species (CREAMS) partial differential
equation solver~\cite{MartinezFerrer2014} is a computational fluid
dynamics application that solves the full Navier-Stokes equations
associated to multi-especies gas mixtures.  CREAMS is \pmf{an
  MPI-only} application written in Fortran and based on the
\pmf{finite difference method}) that is employed for direct numerical
simulations and large eddy simulations.  This application has been
extensively executed on the IBM BlueGene/Q located at IDRIS using up
to $10^5$ MPI processes for grids with approximately $10^9$ points,
see for instance~\cite{MartinezFerrer2017b}.

CREAMS can be run in 1D, 2D and 3D Cartesian grids and supports MPI
domain decomposition in the three spatial directions.  It utilises
sophisticated, \pmf{eighth}-order accurate spatial discretisation
schemes (i.e.\ WENO stencils) requiring four halo elements ($N_h=4$).
A third-order accurate time integration is performed explicitly and in
three consecutive stages, which requires at least three point-to-point
MPI communication per time step.  In the original MPI application,
halos are exchanged in one piece at the end of the integration stage.
It is worth noting that each halo element consists of a
(non-contiguous in memory) list of $N_v=5+N_{\alpha}$ double-precision
real numbers, where $N_v$ is the total number of independent variables
of the problem and $N_{\alpha}$ the number of species (which is about
\pmf{tens} or even hundreds for real-world applications).

The \emph{core} of CREAMS can be considered as a time-loop calling the Runge-Kutta 3 ({\tt rk3}) subroutine that performs the time integration.  This subroutine consists mainly of a loop of three iterations, inside of which there are four differentiated phases: (i) data preparation (lines \rev{2--5}), (ii) WENO stencils (line \rev{6}), (iii) data update (line \rev{7}) and (iv) halo communication (line \rev{8}) as illustrated in Code~\ref{code:rk3-MPI} corresponding to the original, \pmf{MPI-only} application.

\begin{lstfloat}
\begin{lstlisting}[
    caption={CREAMS original, \pmf{MPI-only} Runge-Kutta 3 subroutine.},
    style=customFortran1,
    label={code:rk3-MPI}
  ]
do rk = 1, 3 ! Runge-Kutta 3 loop
   if (rk==1) v0(sx:ex,sy:ey,sz:ez,1:nv) = v(sx:ex,sy:ey,sz:ez,1:nv) ! copy v into v0
   call upd_prims (thd, v, W_i, T, cp, ha)                           ! update primitives
   call upd_boundaries (inp, thd, adi, ctime + cdtime*tk3s, &        ! update boundary
                        x,y,z,dx_i,dy_i,dz_i,T,W_i,cp,ha,v)          ! conditions
   if (ndim >= 1) call euler_LLF_x (thd, T, W_i, cp, ha, v, fhat_x)  ! WENO stencil
   call upd_v (inp,bk3s,ck3s,cdtime,dx_i,dy_i,dz_i,fhat_x,fhat_y,fhat_z,v0,v) ! update v
   call comm_cons_buffers (v) ! halo communications: use MPI subarray datatypes
end do
\end{lstlisting}
\end{lstfloat}

Code~\ref{code:rk3-MPI} also represents the highest level of application code, with computation and communication parts clearly differentiated, where a complete hybrid approach combining MPI domains and OmpSs-2 subdomains can be implemented.  The proposed domain/subdomain methodology seeks to minimise the changes that need to be applied to the original source code, so that it results in a relatively simple refactoring.  This is of great importance considering that the application contains about $10^5$ lines of Fortran source code.  In order to define subdomains, one needs to look firstly at the \pmf{MPI-only} application, i.e.\ at the domains, in which multi-dimensional arrays representing physical variables (such as {\tt v} in Code~\ref{code:rk3-MPI}) are allocated in memory in the following way:
\begin{lstlisting}[style=customFortran2]
  v (sx-nh:ex+nh,sy-nh:ey+nh,sz-nh:ez+nh,1:nv)
\end{lstlisting}
where the (defined in global memory) indexes {\tt nh} correspond to $N_h$, {\tt nv} to $N_v$ and {\tt sx} and {\tt ex} represent the starting and ending MPI domain indexes for the first spatial direction, respectively.  The same applies to $y$- and $z$-directions but, for the sake of conciseness, only the first direction is considered here.  Hence a problem of size {\tt ntx} divided in {\tt nmpix} domains has indexes {\tt sx=1} and {\tt ex=ntx/nmpix} for the first MPI domain, whilst the last one has indexes {\tt sx=1+ntx*(nmpix-1)/nmpix} and {\tt ex=ntx}.  For example, a do-loop over half of each physical (i.e.\ without halos) MPI domain can be simply written as:

\begin{minipage}{\textwidth}
\begin{lstlisting}[style=customFortran2]
do l = 1, nv; do k = ez/2, ez; do j = ey/2, ey; do i = ex/2, ex
  v (i, j, k, l) = 0.0
end do; end do; end do; end do
\end{lstlisting}
\end{minipage}

The next step of the HDOT approach considers that each MPI domain is composed of an arbitrary number $n$ of OmpSs-2 subdomains: for a 1D problem, this is equivalent to divide the range {\tt sx-nh:ex+nh} into $n$ parts.  Typically, the value of $n$ is a multiple (e.g.\ 2, 4, 8\ldots) of the number of cores assigned to shared memory parallel processes.  Furthermore, for 2D and 3D problems subdomain cuts must guarantee that they are always defined in contiguous portions of the original memory layout: this implies cuts in the $y$-direction for 2D problems and in the $z$-direction for 3D problems.

\begin{lstfloat}
\begin{lstlisting}[
    caption={Example of a hybrid loop in CREAMS.},
    style=customFortran1,
    label={code:creams-loop}
  ]
use subdomain                        ! use module subdomain
type (subdomain_type) :: sub         ! declare subdomain derived data type
integer :: s, i0, i1, j0, j1, k0, k1 ! declare local indexes
logical :: dummy                     ! declare dummy boolean variable
call subdomain_ini (grainsize, sub)  ! initialise sub: get the number of subdomains
do s = 1, sub % n                    ! loop over all subdomains
   call subdomain_set (s, sub)       ! get boundaries (indexes) for current subdomain
   call subdomain_idx (sub  ,ex/2,ex,ey/2,ey,ez/2,ez, & ! convert global MPI indexes
                       dummy,i0  ,i1,j0  ,j1,k0  ,k1)   ! into local subdomain indexes
   if (.not. dummy) then ! run a task (operation) for each "useful" subdomain
      !$OSS TASK FIRSTPRIVATE (i0,i1,j0,j1,k0,k1,nv) PRIVATE (i,j,k,l) SHARED (v)
      do l = 1, nv; do k = k0, k1; do j = j0, j1; do i = i0, i1
         v (i, j, k, l) = 0.0
      end do; end do; end do; end do
      !$OSS END TASK
   end if; end do
\end{lstlisting}
\end{lstfloat}

In order to effectively incorporate subdomains in the source code, it becomes necessary to define a Fortran module containing a derived data type representing the subdomain, e.g.\ total number of subdomains, domain ID and its boundaries (indexes), as well as functions to manipulate the data type.  This approach allows writing a hybrid version of the previous nested loop as shown in Code~\ref{code:creams-loop}.  This self-explanatory code only depends on the input variable {\tt grainsize} corresponding to the subdomain partition size defined by the user.  Moreover, global indexes {\tt sx} and {\tt ex} are replaced by local indexes {\tt i0} and {\tt i1} via the function {\tt subdomain\_idx}.  This change from global to local indexes is necessary to work with subdomains and constitutes indeed the major code refactoring.

Nevertheless, the changes to be made are pretty straightforward and only require defining local integers and passing an additional argument ({\tt sub}) as a first-private copy in the corresponding subroutines, thus leaving intact the original memory layout of the application and ensuring a thread-safe execution.

Moreover, when using subdomains the new local indexes must always refer to absolute coordinates.  For instance, defining a symmetry-type boundary condition at the rightmost point {\tt ntx} using global, \emph{relative} indexes:

\begin{minipage}{\textwidth}
\begin{lstlisting}[style=customFortran2]
do l = 1, nv; do k = sz, ez; do j = sy, ey; do i = 1, nh
  v (ntx+i, j, k, l) = v (ntx-i, j, k, l)
end do; end do; end do; end do
\end{lstlisting}
\end{minipage}

\noindent must be replaced by local, \emph{absolute} indexes:

\begin{minipage}{\textwidth}
\begin{lstlisting}[style=customFortran2]
do l = 1, nv; do k = k0, k1; do j = j0, j1; do i = i0, i1 ! i0=ntx+1
  v (i, j, k, l) = v (ntx+ntx-i, j, k, l)                 ! i1=ntx+nh
end do; end do; end do; end do
\end{lstlisting}
\end{minipage}

\noindent for subdomains indexes to be properly defined, which sometimes might be a bit counterintuitive: it is indeed much easier to perform a loop from {\tt 1} to {\tt nh} and then count backwards and forwards from the symmetry point {\tt ntx} as in the first example corresponding to the original, \pmf{MPI-only} application.

\begin{lstfloat}
\begin{lstlisting}[
    caption={CREAMS hybrid Runge-Kutta 3 subroutine (OmpSs-2 pragmas heavily simplified).},
    style=customFortran1,
    label={code:rk3-hybrid}
  ]
call subdomain_ini (grainsize, sub)
do rk = 1, 3; do s = 1, sub % n
   call subdomain_set (s, sub)
   !$OSS TASK FIRSTPRIVATE (sub) PRIVATE (i0, i1) OUT (v0) INOUT (T, W_i, cp, ha, v)
   call subdomain_idx (sub, sx, ex, sy, ey, sz, ez, dummy, i0, i1, j0, j1, k0, k1)
   if (rk==1 .and. .not. dummy) v0(i0:i1,j0:j1,k0:k1,1:nv) = v(i0:i1,j0:j1,k0:k1,1:nv)
   call upd_prims (sub, thd, v, W_i, T, cp, ha)
   call upd_boundaries (sub, inp, thd, adi, ctime + cdtime*tk3s,      &
                        x, y, z, dx_i, dy_i, dz_i, T, W_i, cp, ha, v)
   !$OSS END TASK
   call subdomain_idx (sub,sx-1,ex,sy-1,ey,sz-1,ez,dummy,i0,i1,j0,j1,k0,k1)
   if (.not. dummy .and. ndim >=1) then
      !$OSS TASK FIRSTPRIVATE (sub) IN (T, W_i, cp, ha, v) OUT (fhat_x)
      call euler_LLF_x (sub, thd, T, W_i, cp, ha, v, fhat_x)
      !$OSS END TASK
   end if
   call subdomain_idx (sub,sx,ex,sy,ey,sz,ez,dummy,i0,i1,j0,j1,k0,k1)
   if (.not. dummy) then
      !$OSS TASK FIRSTPRIVATE (sub) IN (T, W_i, fhat_x, fhat_y, fhat_z, v0) INOUT (v)
      call upd_v (sub,inp,bk3s,ck3s,cdtime,dx_i,dy_i,dz_i,fhat_x,fhat_y,fhat_z,v0,v)
      !$OSS END TASK
   end if; end do
   call fill_cons_buffers  (v) ! fill custom send buffers before communications
   call comm_cons_buffers  (v) ! communicate data from send buffers to recv buffers
   call empty_cons_buffers (v) ! empty custom recv buffers after communications
end do
\end{lstlisting}
\end{lstfloat}

\begin{figure}[tp]
  \begin{minipage}[c]{\textwidth}
    \begin{center}
      \hspace*{\fill}
      \subfloat[]{
        \includegraphics[height=6.5cm]{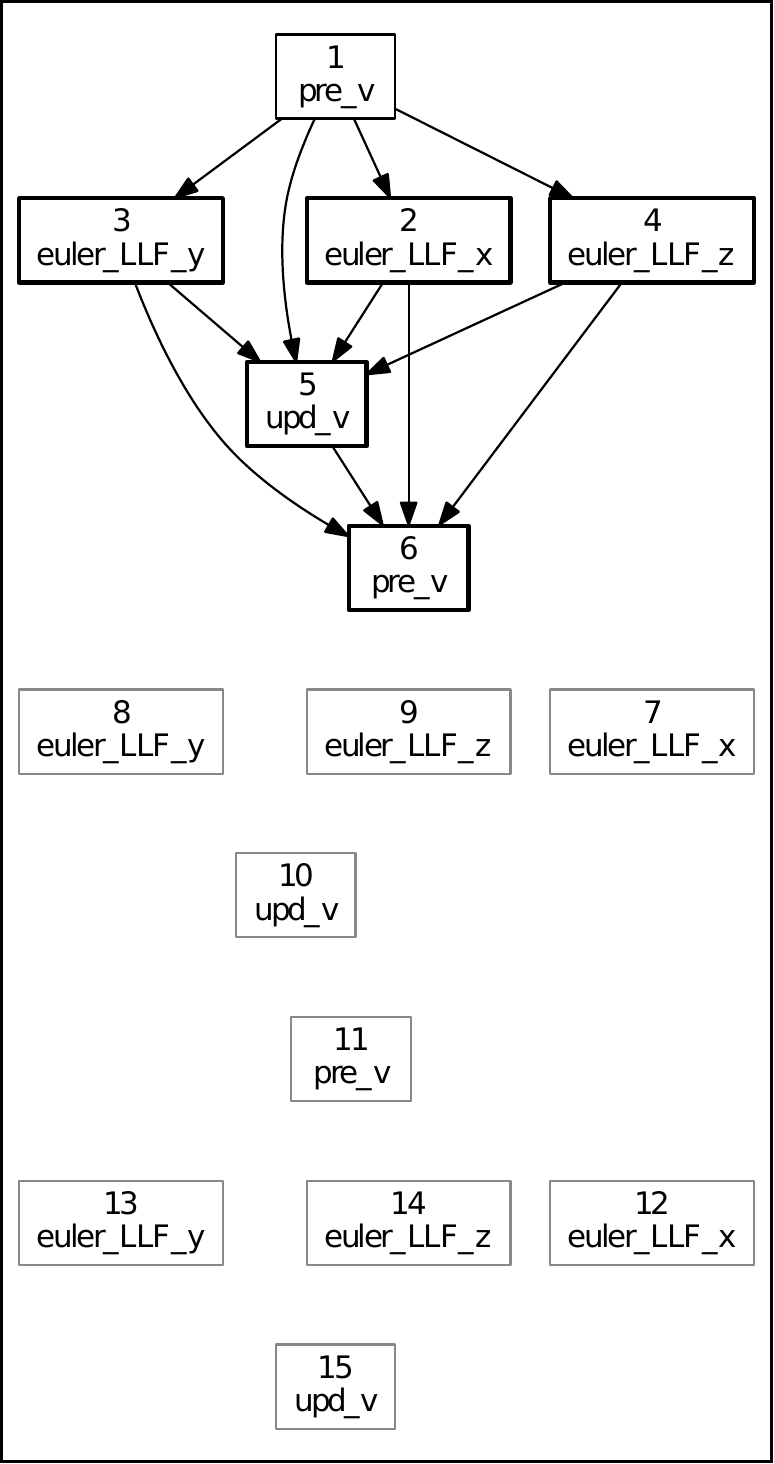}\label{fig:graph-1sub-start}
      }
      \hfill
      \subfloat[]{
        \includegraphics[height=6.5cm]{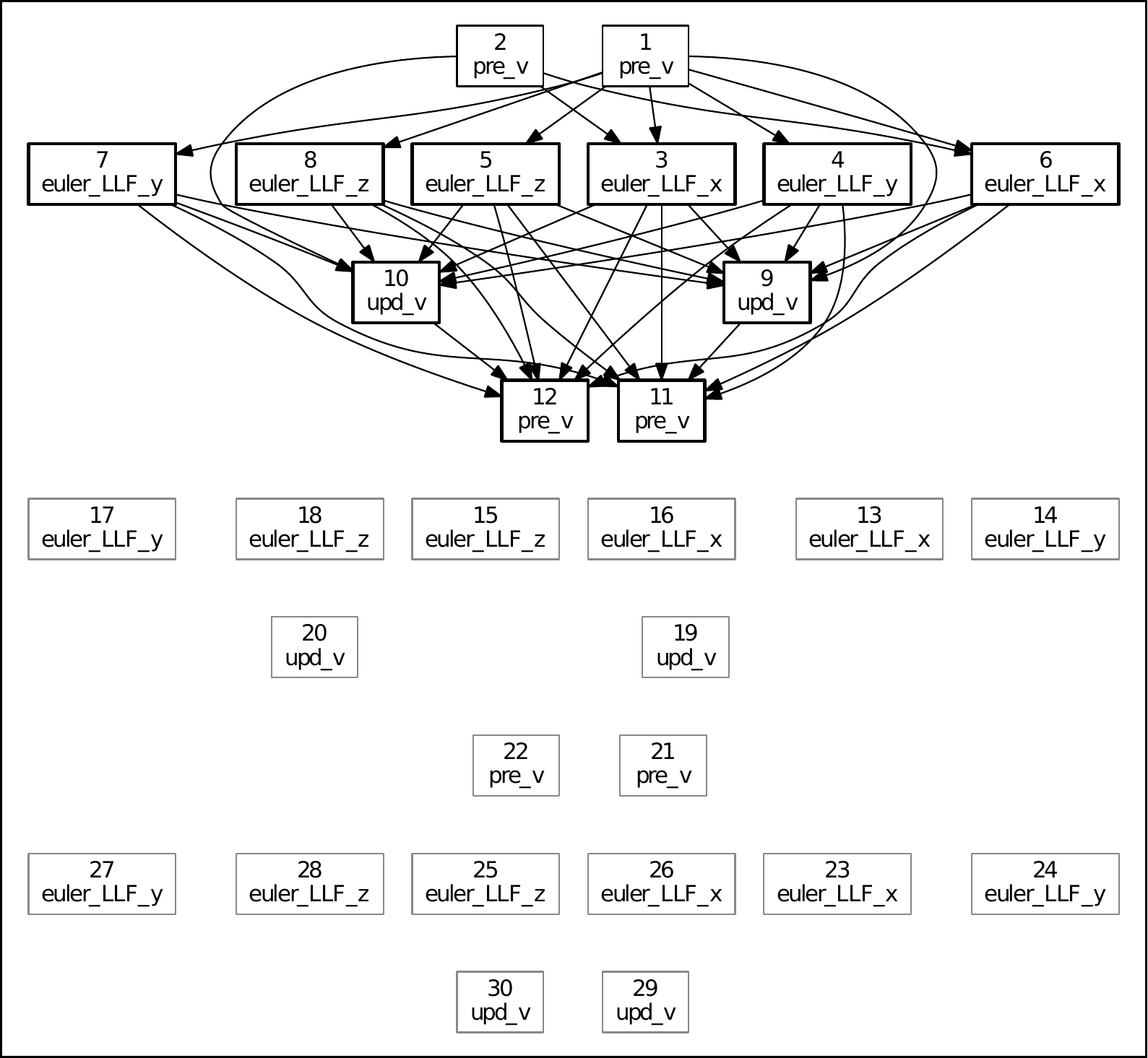}\label{fig:graph-2sub-start}
      }
      \hspace*{\fill}
    \end{center}
  \end{minipage}
  \caption{CREAMS dependencies graph corresponding to the first stage of the Runge-Kutta time integration (the other two are just a repetition) for (a) one OmpSs-2 subdomain and (b) two OmpSs-2 subdomains;  WENO stencils ({\tt euler\_LLF\_*}) can run in parallel from each other and {\tt pre\_v} and {\tt upd\_v} refer to data preparation and data update tasks, respectively.}
  \label{fig:creams-graphs}
\end{figure}

\begin{figure}[tp]
  \begin{center}
    \includegraphics[width=0.75\textwidth]{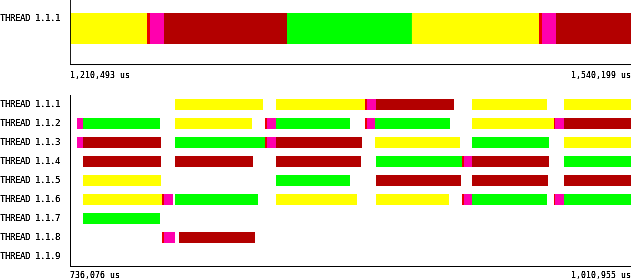}
  \end{center}
  \caption{CREAMS Paraver trace for one OmpSs-2 subdomain forced to be
    executed with \rev{one} thread (top) and two OmpSs-2 subdomains
    \rev{without thread restrictions} (bottom); tasks and colours are
    as follows: {\tt pre\_v} (pink), {\tt euler\_LLF\_x} (brown), {\tt
      euler\_LLF\_y} (green), {\tt euler\_LLF\_z} (yellow) and {\tt
      upd\_v} (red).}
  \label{fig:creams-traces}
\end{figure}

On the other hand, it is important to highlight that the creation of
subdomains/tasks is not performed at low-level loops such as the one
shown in Code~\ref{code:creams-loop}, but at the highest possible
level of the application code: the {\tt rk3} subroutine \pmf{itself}.
As a result, the hybrid version of CREAMS, see
Code~\ref{code:rk3-hybrid}, looks quite similar to the original
\pmf{MPI-only} implementation.  Leaving out pragma annotations, which
are heavily simplified here due to page restriction, the application
core becomes \emph{completely} hybrid using only three subdomain/task
groups (lines \rev{4--10}, \rev{13--15} and \rev{19--21} of
Code~\ref{code:rk3-hybrid}) for computations as illustrated in
Figure~\ref{fig:creams-graphs}.  From this figure it can be readily
seen that the WENO stencils can be executed simultaneously as they are
placed in the same horizontal time line.  For one subdomain and 3D
problems there are three simultaneous stencil executions (one in each
spatial dimension), and this number doubles for two subdomains.
Paraver traces depicted in Figure~\ref{fig:creams-traces} give a
better understanding of the OmpSs-2 real-time execution.  The
single-threaded CREAMS execution of Figure~\ref{fig:creams-traces}
reveals that WENO stencils are indeed the most computationally
demanding tasks, which account for more than 95\% of the total
\rev{runtime on} average.  \rev{However, w}ith two subdomains, there
are always six busy threads computing WENO stencils in either
direction and, when data needs to be prepared or updated, this number
reduces to the \pmf{actual} number of subdomains, following the graph
patterns exposed in Figure~\ref{fig:creams-graphs}.


The source code containing \pmf{MPI communication} also needs to be
adapted to work with subdomains.  In the original, \pmf{MPI-only}
application, MPI subarray \pmf{datatypes} are used to duplicate halos,
whilst communication (line \rev{8} of Code~\ref{code:rk3-MPI}) simply
\pmf{consist} of calling all the \pmf{non-blocking sends ({\tt
    MPI\_ISEND}) and receives ({\tt MPI\_IRECV})} functions with a
final \pmf{blocking} call to {\tt MPI\_WAITALL} that synchronises the
entire data exchange.  For the hybrid version, instead of using MPI
subarray \pmf{datatypes}, we define custom (contiguous in memory)
send/recv buffers that are filled (line \rev{23} of
Code~\ref{code:rk3-hybrid}) and emptied (line \rev{25} of
Code~\ref{code:rk3-hybrid}), respectively, between an MPI
communication to ensure that the data is available before sending it
and also available after receiving it and used by another task.

\begin{lstfloat}
\begin{lstlisting}[
    caption={CREAMS hybrid communication with East neighbours (OmpSs-2 pragmas heavily simplified).},
    style=customFortran1,
    label={code:commx-hybrid}
  ]
call subdomain_ini (grainsize, sub); countSendE = 1; countRecvE = 1
do s = 1, sub % n ! a single subdomain loop to send/recv communication buffers
   call subdomain_set (s, sub)
   call subdomain_idx (sub,ex-ng+1,ex,sy,ey,sz,ez,dummy,i0,i1,j0,j1,k0,k1)
   if (.not. dummy .and. neigh (E) /= MPI_PROC_NULL) then ! send to East
      bsize = nv*(i1-i0+1)*(j1-j0+1)*(k1-k0+1)
      !$OSS TASK FIRSTPRIVATE (i0, j0, k0, bsize, countSendE) IN (consBuffSendE)
      call MPI_ISEND(consBuffSendE(1,i0,j0,k0),bsize,MPI_DOUBLE_PRECISION,neigh(E), &
                     1000 + countSendE, comm3d, reqSendE (countSendE), mpicode)
      call TAMPI_IWAIT(reqSendE (countSendE), MPI_STATUS_IGNORE, mpicode)
      !$OSS END TASK
      countSendE = countSendE + 1
   end if
   call subdomain_idx (sub,ex+1,ex+ng,sy,ey,sz,ez,dummy,i0,i1,j0,j1,k0,k1)
   if (.not. dummy .and. neigh (E) /= MPI_PROC_NULL) then ! receive from East
      bsize = nv*(i1-i0+1)*(j1-j0+1)*(k1-k0+1)
      !$OSS TASK FIRSTPRIVATE (i0, j0, k0, bsize, countRecvE) OUT (consBuffRecvE)
      call MPI_IRECV(consBuffRecvE(1,i0,j0,k0),bsize,MPI_DOUBLE_PRECISION,neigh(E), &
                     2000 + countRecvE, comm3d, reqRecvE (countRecvE), mpicode)
      call TAMPI_IWAIT(reqRecvE (countRecvE), MPI_STATUS_IGNORE, mpicode)  
      !$OSS END TASK
      countRecvE = countRecvE + 1
   end if; end do

\end{lstlisting}
\end{lstfloat}

The actual MPI communication is performed by the {\tt
  comm\_cons\_buffers} subroutine (line \rev{24} of
Code~\ref{code:rk3-hybrid}) and takes place inside subdomain tasks
using the TAMPI library.  As an example, the communication involving
the East neighbour, i.e.\ the neighbour following the positive
$x$-direction, \rev{is} shown in Code~\ref{code:commx-hybrid}.  The
code structure remains similar to the one used for computation tasks.
\pmf{Herein, the TAMPI subroutine {\tt TAMPI\_IWAIT} guarantees that
  the corresponding tasks do not get blocked waiting to receive the
  buffer and, instead, return immediately.}  This way MPI waits are
completely substituted by {\tt IN} and {\tt OUT} pragma clauses taking
place during the filling and emptying of communication buffers (lines
\rev{23} and \rev{25}, respectively, of Code~\ref{code:rk3-hybrid})
and the generated dependencies are managed by OmpSs-2 as if they were
another part of the computation.  \pmf{Note also that
  Code~\ref{code:rk3-hybrid}, corresponding to the hybrid version of
  CREAMS, will only work with a version of the MPI library with
  multi-threading support due to the use of TAMPI, whilst the
  MPI-only version is executed with a single MPI thread.}

\begin{figure}[tp]
  \begin{minipage}[c]{\textwidth}
    \begin{center}
      \hspace*{\fill}
      \subfloat[]{
        \includegraphics[height=6.5cm]{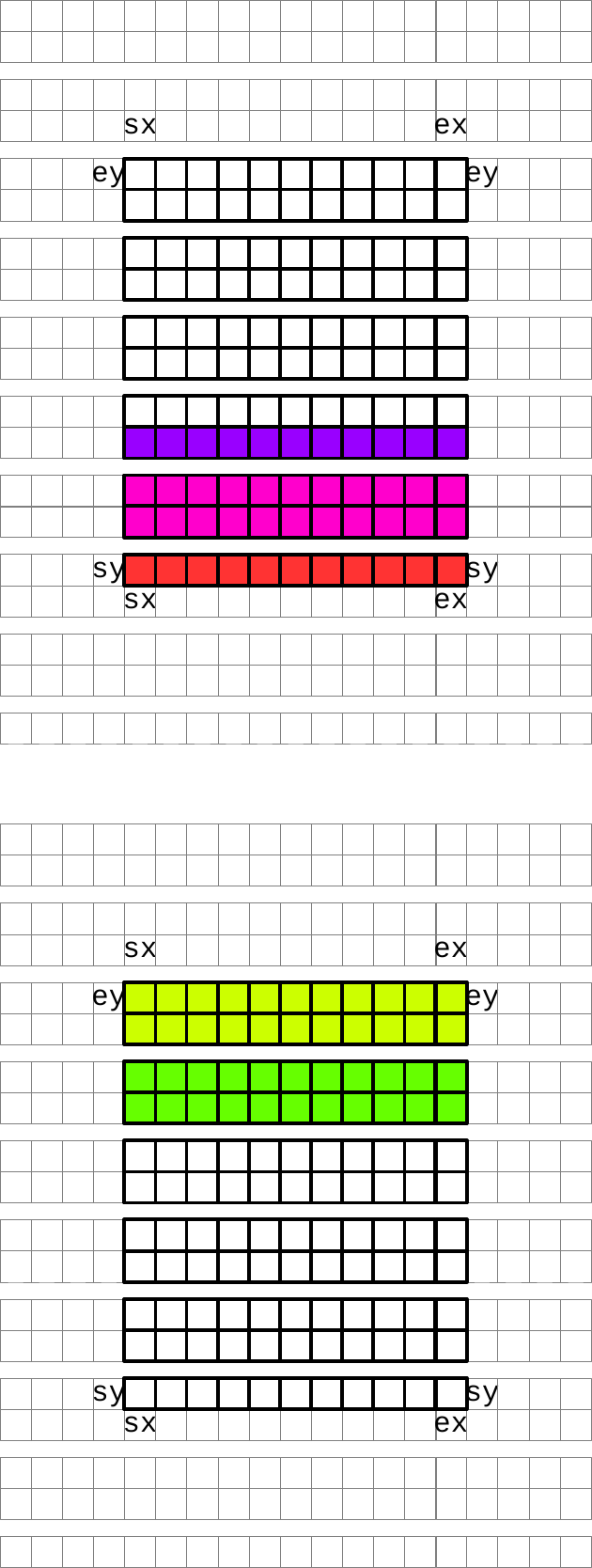}\label{fig:commParBad}
      }
      \hfill
      \subfloat[]{
        \includegraphics[height=6.5cm]{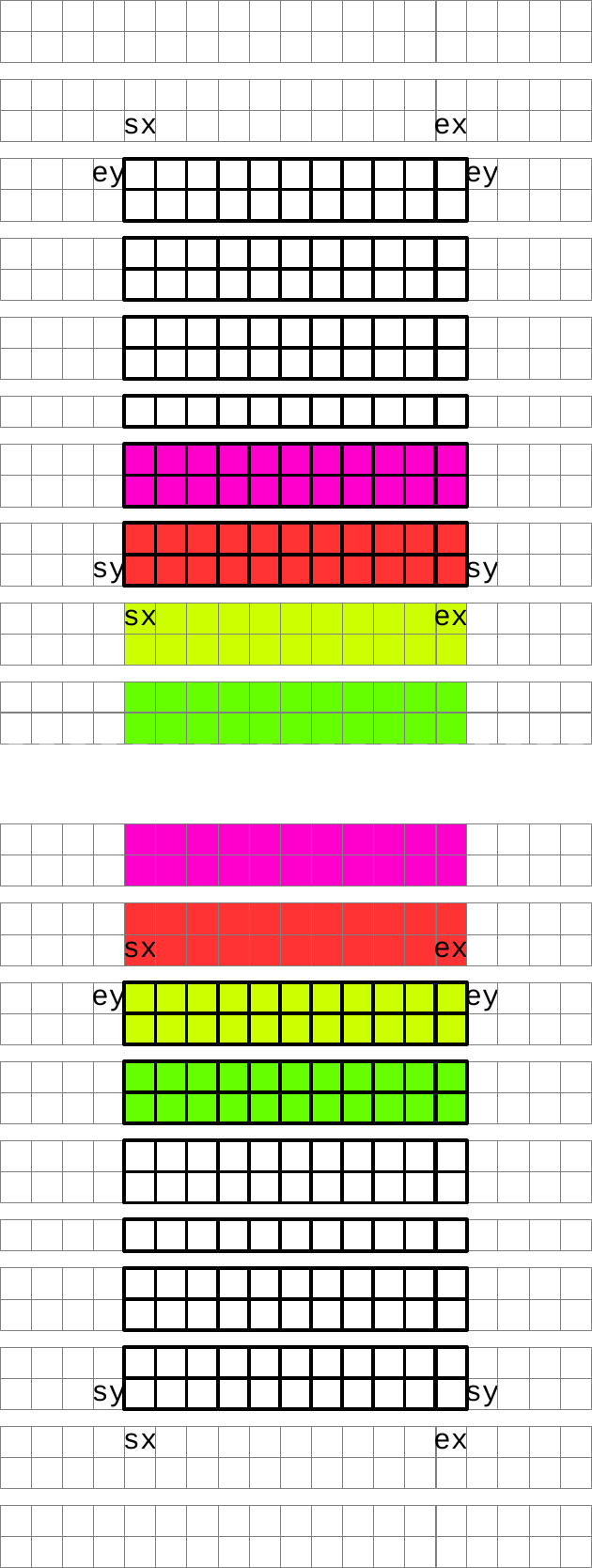}\label{fig:commParGood}
      }
      \hfill
      \subfloat[]{
        \includegraphics[height=6.5cm]{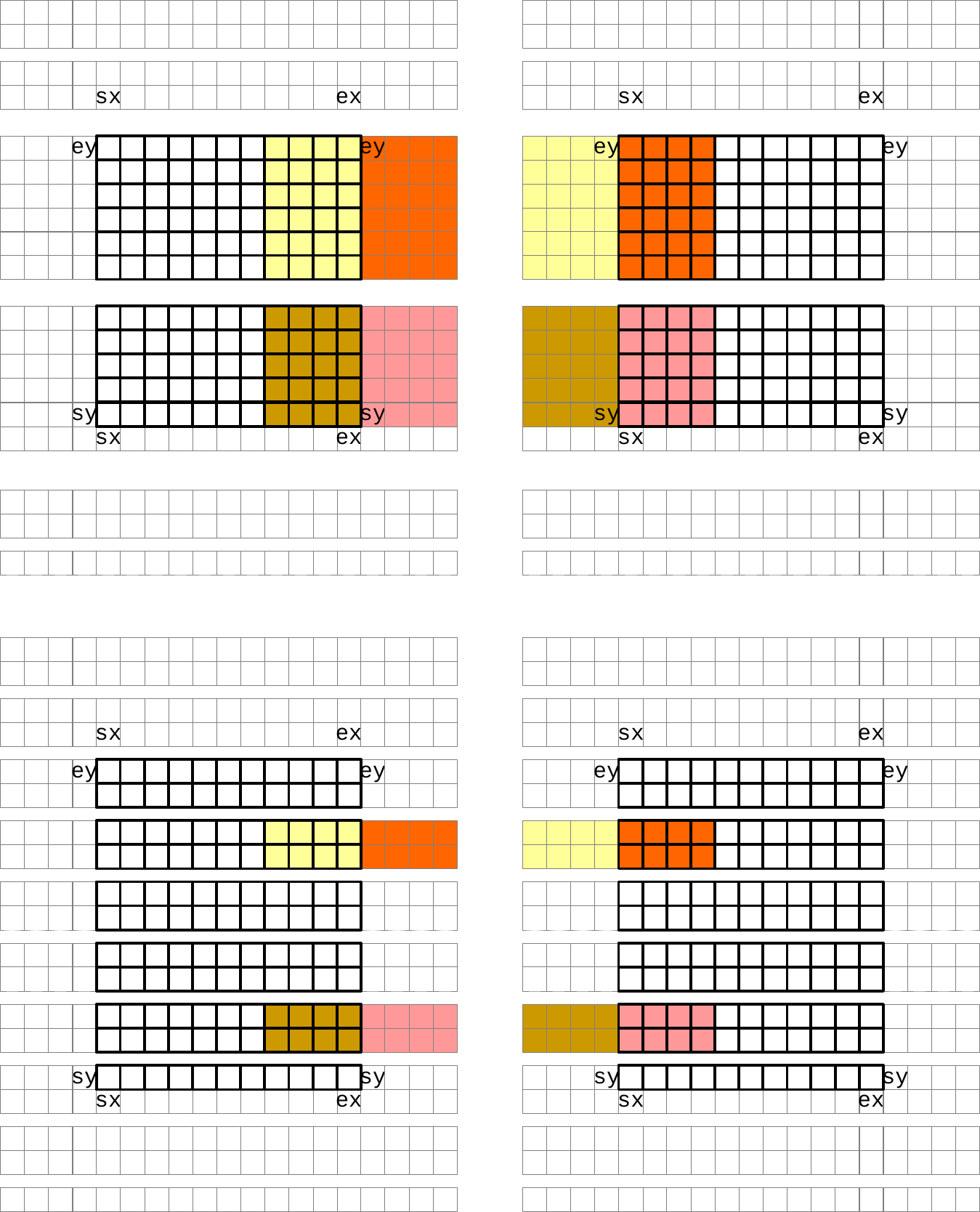}\label{fig:commOrt}
      }
      \hspace*{\fill}
    \end{center}
  \end{minipage}
  \caption{CREAMS possible subdomain decompositions yielding to (a) invalid parallel MPI communication, (b) valid parallel MPI communication and (c) always valid orthogonal MPI communication.}
  \label{fig:creams-comms}
\end{figure}

\pmf{It should be noted that it is also possible to embed
  communication inside tasks without relying upon the TAMPI library
  (not shown here for the sake of conciseness).  This procedure is
  similar to the one shown in Code~\ref{code:commx-hybrid} except
  that: (i) two subdomain loops are required (instead of a single
  one), the first loop \emph{only to send} the communication buffers,
  and the second loop \emph{only to receive} the communication buffers
  and where {\tt TAMPI\_IWAIT} is replaced by {\tt MPI\_WAIT}; (ii)
  \rev{it must be granted that} all the send buffers \rev{are} sent
  before start receiving any recv buffer.  It can be readily seen that
  this alternative approach requires the programmer to ensure that no
  MPI dead locks occur by enforcing the order in which
  tasks/communication must be executed \rev{via blocking directives
    which ultimately yield an innefficient usage of MPI.}}

There are two important aspects regarding MPI communication with
subdomains that are worth discussing.  The first one deals with the
communication in the direction that is \pmf{\emph{parallel}} to
subdomains, e.g.\ the $y$-direction in a 2D problem, see
Figures~\ref{fig:creams-comms}a--b.  One needs to take into account
that a point-to-point send/recv communication \rev{is} always
asymmetric with respect to the communication edges of the two involved
MPI domains.  In Figure~\ref{fig:creams-comms}a, the proposed
subdomain decomposition is not asymmetric and hence there is not a
valid correspondence between the three tasks from the top domain and
the two from the bottom \pmf{domain}.  In CREAMS the number of halo
\rev{elements} is $N_h=4$ and hence the minimum allowed {\tt
  grainsize} values are 1, 2 (see Figure~\ref{fig:creams-comms}b) or 4
to guarantee asymmetry.  On the other hand, an MPI communication
\pmf{\emph{orthogonal}} to subdomain cuts does not have any
constraints in terms of subdomain decomposition, see
Figure~\ref{fig:creams-comms}c.

The second aspect concerns the actual implementation of MPI
communication within subdomains.  The proposed implementation needs
three subroutines (lines \rev{23--25} of Code~\ref{code:rk3-hybrid})
and therefore implies generating three groups of additional tasks and
thus potentially more overhead for the OmpSs-2 runtime.  A more
optimised implementation could be achieved by embedding these calls
inside the computation tasks, which would require modifying the pragma
clauses accordingly.  This implementation will be studied in a later
development phase of the application as the scalability results
obtained using the proposed approach already look quite promising.

Finally, Table~\ref{tab:CREAMS-scalability} shows CREAMS scalability results obtained with the \pmf{MPI-only} and hybrid versions of the application.  These results correspond to the Sod tube benchmark, which is commonly employed to test the robustness and accuracy of numerical schemes for compressible flows~\cite{MartinezFerrer2014} and is also suitable for measuring scalability performance~\cite{Macia2018}.  The dimensions of the computational domain are $N_x \times N_y \times N_z = 20 \times 20 \times 7000$, resulting in a total of 2.8 million grid points, and runtimes values are measured after 1000 complete time steps using up to 16 nodes of the MareNostrum 4 supercomputer.  Hybrid computations are set up in such a way that they only use 2 MPI domains per node (one MPI domain per socket), thus reducing the total amount of MPI messages exchanged between ranks, and each MPI domain have all the available 28 cores per socket to carry out subdomain/task computations.  Even when using a single node, the hybrid version is 2.58\% faster than the \pmf{MPI-only} version.  As the number of nodes increases, this gain becomes non-negligible and, in this particular benchmark, reaches a maximum value of 13.33\%.  This difference is mainly due to the increasing amount of communication among the 768 MPI processes (16 nodes) present in the \pmf{MPI-only} version compared to the only 32 MPI processes required by the hybrid version.

The above results confirm the advantages of hybrid parallel programming over the classical MPI decomposition.  \pmf{It is worth mentioning that, on the one hand, the communication pattern of the original MPI-only version does not allow overlapping communication with computation thus yielding suboptimal performance.  On the other hand, the results of the hybrid version of CREAMS correspond to early stages of development} and, consequently, better figures should be expected with the implementation of array reductions and tasks loops that are not currently fully supported in the Fortran version of OmpSs-2.  Similarly, better scalability is expected with the integration of MPI communication inside computation tasks \pmf{and by reducing the number of MPI processes to only one per node with an interleaved memory NUMA policy between sockets.}

%
\begin{table}[btp]
  \footnotesize
  \centering
  \caption{CREAMS Sod tube parallel scalability results on MareNostrum 4; indexes M and H refer to the \pmf{MPI-only} and hybrid (MPI+OmpSs-2) CREAMS application, respectively.  SI units.}
  \begin{tabular}{cS[table-format=3.3]S[table-format=3.3]S[table-format=2.2]S[table-format=2.2]S[table-format=3.2]S[table-format=3.2]S[table-format=1.2]}
    \toprule
    Nodes &
    $\mathrm{Runtime_{M}}$ & $\mathrm{Runtime_{H}}$ &
    $\mathrm{S._{M}}$ & $\mathrm{S._{H}}$ &
    $\mathrm{Perf._{M}}$ & $\mathrm{Perf._{H}}$ &
    $\mathrm{Gain_{M \to H}}$ \\
    \midrule
    \phantom{0}1  & 962.681 & 937.886 & 1     & 1     & 100.00\% & 100.00\% & +2.58\% \\
    \phantom{0}2  & 485.431 & 470.240 & 1.98  & 1.99  &  99.16\% &  99.72\% & +3.13\% \\
    \phantom{0}4  & 250.200 & 235.342 & 3.85  & 3.98  &  96.19\% &  99.63\% & +5.94\% \\
    \phantom{0}8  & 131.880 & 118.728 & 7.30  & 7.90  &  91.25\% &  98.74\% & +9.97\% \\
              16  &  69.705 &  60.415 & 13.81 & 15.52 &  86.32\% &  97.03\% & +13.33\% \\
    \bottomrule
  \end{tabular}
  \label{tab:CREAMS-scalability}
\end{table}
%

\pmf{
  Finally, we would like to summarise the necessary code changes required to apply the HDOT methodology on an MPI-only, complex application composed of hundreds of thousands of lines of source code.  Firstly, to avoid rewriting the whole application from scratch, it is necessary to build a new module or class that helps handling the original data layout (where indexes are global) via subdomain/tasks (where indexes become local).  This still implies some refactoring as shown in Code~\ref{code:creams-loop}.  Secondly, it is necessary to identify the highest possible level of application code to define subdomains/tasks in order to: (i) reduce the refactoring effort to the strict minimum, (ii) also reduce the total amount of tasks per execution (thus yielding better performance) and (iii) avoid leaving parts of the code serialised.  By combining these two steps, subdomains/tasks will cover vast regions of the application code and basically consist of calls to other functions instead of small kernel loops as shown in Code~\ref{code:rk3-hybrid}.  Note also that OmpSs-2 pragma annotations will only be inserted at this level.  Thirdly, MPI communication must also be made compatible with the specification of subdomain/tasks and hence require custom send/recv buffers if this was not originally the case (e.g.\ MPI subarray datatypes originally used CREAMS, which greatly simplify the use of communication buffers, cannot longer be utilised).  Moreover, subdomain decomposition must be consistent with communication patterns as shown in Figure~\ref{fig:creams-comms}.  Finally, there are multiple ways to perform MPI communication with tasks/subdomains, and the best implementation seeks to overlap it with computation (avoiding potential dead locks).  The approach retained here achieves that with the aid of the TAMPI library, which greatly simplifies the treatment of dependencies between OmpSs-2 tasks and avoids MPI deadlocks.
}

%% file: tex/conclusion.tex
\section{Conclusion and future work}
\label{sec:conclusion}

In this work, we have presented a methodology to develop hybrid
applications that overcome the main issues of \pmf{MPI-only} and
traditional hybrid MPI+OpenMP applications.  The main point of this
methodology is to apply the original MPI domain decomposition strategy
as implemented on process-level to task-level. This way each part of
the global domain assigned to \pmf{an} MPI rank is divided in
subdomains that will be processed by OmpSs-2 tasks. This paper
describes how typical computation and communication patterns have to
be modified to obtain the best performance while maximising code
reuse. The hierarchical domain over-decomposition with tasking (HDOT)
methodology also exploits synergies between MPI and OmpSs-2: on the
one hand, it reuses the original MPI parallelisation strategy to
expose coarse-grained parallelism inside a node; on the other hand,
the OmpSs-2 data-flow execution based on fine-grained dependencies is
leveraged to provide fine-grained, internode synchronisation.  HDOT
relies on the task-aware MPI (TAMPI) interoperability library as well
as advanced support of task nesting and fine-grained dependencies
provided by OmpSs-2 to minimise the structural changes required to
develop a performant hybrid version.  \rev{Although it would be
  possible to achieve similar performance improvements by (i) using
  the current features provided by MPI such as shared-memory, one-side
  communication, (ii) developing a custom user runtime on top of MPI
  and (iii) implementing a custom interoperatibility library.  We
  think that this solution will be more complex, harder to use and it
  will require significant changes in the application structure.}

\pmf{\rev{HDOT} has been applied to two benchmarks (Heat2D and
  HPCCG) and one large application (CREAMS) and the performance
  results have shown a clear gain in performance over other
  traditional approaches relying on either MPI exclusively or
  MPI+OpenMP.  Nevertheless,} \xt{ additional testing with more
  applications will be required to further refine it. Specifically, we
  plan to apply this methodology to other well-known benchmarks (e.g.\
  LULESH~\cite{LULESH}) and to a particle-in-cell based method
  (similar to iPIC3D~\cite{ipic3d}) to improve our current
  discussion. The main goal of these future studies will be focused on
  finding new patterns that could be included in the current
  methodology, but also on looking for boundaries of this approach.}

In addition, the HDOT methodology is applicable to other message
passing APIs such as GASPI making use of advanced features of the
OmpSs-2 runtime system.  \xt{We are interested in presenting this
  approach as well as the discussed runtime features to the OpenMP
  architecture review board in order to promote their adoption for
  hybrid programming. We will also explore the oportunities to extend
  the current specification of the OpenMP {\tt detach} clause in order
  to be more flexible with respect to the number of non-blocking
  services involved in the task finalisation.}